\documentclass[11pt,oneside]{article}
\pdfoutput=1
\usepackage{setspace,graphicx,amssymb,amsmath,latexsym,amsfonts,amscd,amsthm,multirow,ctable,mathdots,caption,array,diagbox,mathtools}
\usepackage{tabularx,cite,mathrsfs}
\usepackage{authblk}
\usepackage{color}
\usepackage{tabu}

\usepackage[margin=2cm]{caption}

\usepackage{fullpage}
\usepackage{stmaryrd}
\usepackage{rotating}

\usepackage{hyperref}

\newcommand{\bea}{\begin{eqnarray}}
\newcommand{\eea}{\end{eqnarray}}
\newcommand{\bee}{\begin{eqnarray*}}
\newcommand{\eee}{\end{eqnarray*}}
\newcommand{\al}{\begin{align*}}
\newcommand{\eal}{\end{align*}}
\newcommand{\be}{\begin{equation}}
\newcommand{\ee}{\end{equation}}
\newcommand{\bem}{\begin{pmatrix}}
\newcommand{\eem}{\end{pmatrix}}

\newcommand{\mc}{\mathcal}
\newcommand{\bb}{\mathbb}

\def\a{\alpha}

\def\c{\gamma}

\def\t{\tau}
\def\th{\theta}

\def\Tr{\text{Tr}}

\def\el{\in}
\def\({\left(}
\def\){\right)}
\def\nn{\nonumber}
\numberwithin{equation}{section}

\begin{document}

\setstretch{1.4}

\title{\vspace{-65pt}
\vspace{20pt}
    \textsc{
    Symmetries of the refined D1/D5 BPS spectrum
    }
}

\author[1]{Nathan Benjamin\thanks{nathansbenjamin@gmail.com}
}

\author[2]{Sarah M. Harrison\thanks{sarharr@physics.harvard.edu}
}

\affil[1]{Stanford Institute for Theoretical Physics, Department of Physics\\
Stanford University, Stanford, CA 94305, USA}

\affil[2]{Center for the Fundamental Laws of Nature, 
Harvard University, Cambridge, MA 02138, USA}

\date{}

\maketitle

\vspace{-1em}

\centerline{\emph{This paper is dedicated to the $23^{rd}$ President of the United States, Benjamin Harrison.}}

\abstract{
We examine the large $N$ 1/4--BPS spectrum of the symmetric orbifold CFT Sym$^N(M)$ deformed to the supergravity point in moduli space for $M= K3$ and $T^4$. We consider refinement under both left- and right-moving $SU(2)_R$ symmetries of the superconformal algebra, and decompose the spectrum into characters of the algebra. We find that at large $N$ the character decomposition satisfies an unusual property, in which the degeneracy only depends on a certain linear combination of left-- and right--moving quantum numbers, suggesting deeper symmetry structure. 
Furthermore, we consider the action of discrete symmetry groups on these degeneracies, where certain subgroups of the Conway group are known to play a role. We also comment on the potential for larger discrete symmetry groups to appear in the large $N$ limit.
}

\clearpage

\tableofcontents

\section{Introduction}
\label{sec:intro}
The study of BPS spectra of string compactifications with a large number of supercharges has furnished a laboratory for a number of exact results in a theory of quantum gravity, including a microscopic description of black hole entropy \cite{Strominger:1996sh} and examples of the AdS/CFT correspondence \cite{Maldacena:1997re}. If one considers the D1/D5 system compactified on $M \times S^1$, where $M= K3 $ or $T^4$, the worldsheet superconformal field theory (SCFT) of the resulting string in spacetime lies in the moduli space of the symmetric product conformal field theory Sym$^N(M)$ where, in the case of $Q_1$ D1-branes wrapping $S^1$ and $Q_5$ D5-branes wrapping $M \times S^1$, $N=Q_1Q_5$. Furthermore, in the case of $K3$, the elliptic genus (EG) of the $\mathcal N=(4,4)$ worldsheet SCFT, defined as,\footnote{Below and throughout the text, we make use of the definitions $q= e^{2\pi i \tau}$, $y=e^{2\pi i z}$, and $u=e^{2\pi i \nu}$.}
\be
Z_{\text{EG}}(\tau, z) = \Tr_{\text{RR}}\((-1)^{F_L+F_R}q^{L_0-\frac{c}{24}}y^{J_0} \),\footnote{Throughout the text we use $J_0$ to denote the Cartan of the $\mathcal N=4$ $SU(2)$.}
\label{eq:EG}
\ee
which counts spacetime 1/4--BPS states, was shown to reproduce the Bekenstein-Hawking entropy of the corresponding five-dimensional black hole \cite{Strominger:1996sh}. A similar accounting of spacetime 1/8--BPS states for $M=T^4$ coming from a modified worldsheet index was shown to govern the entropy of $\mathcal N=8$ black holes in \cite{Maldacena:1999bp}.

In \cite{Kachru:2016igs}, a new quantity called the Hodge elliptic genus (HEG) was defined, and was proposed to compute degeneracies of 1/4--BPS states in string theory on $K3 \times S^1$, flavored under both the left- and right-moving $SU(2)$ angular momentum quantum numbers of the $SO(4)$ little group in five dimensions, thus furnishing a more refined count of BPS states. It is clear from the definition of the HEG, 
\be
Z_{\text{HEG}}(\tau, z, \nu) = \Tr_{\text{RR}}\((-1)^{F_L+F_R}q^{L_0-\frac{c}{24}}y^{J_0} u^{\bar{J_0}} \) \bigg|_{\bar h = \frac c{24} }, 
\label{eq:HEG}
\ee
that unlike the EG, the HEG is not an index in the sense that it depends on the point in CFT moduli space at which it is computed. That is, by tracing over right-moving Ramond ground states graded by the $U(1)$ charge that is part of the right-moving $\mathcal N=2$ superconformal algebra, this quantity can jump as one moves in CFT moduli space. Furthermore, the EG is known to have nice modular properties whenever the CFT has compact target space; i.e., for an $\mathcal N=(2,2)$ superconformal field theory with central charge $c=6m$, the EG is known to be a weak Jacobi form of weight zero and index $m$. However, the HEG in general is not known to enjoy such nice modular properties.

Other recent work has employed the HEG to study the growth of refined BPS states at the supergravity point in the moduli space of Sym$^N(M)$ for $M=K3$ and  $T^4$ at large $N$ \cite{Benjamin:2016pil}, as well as the refined spectrum of the D1/D5 system on $T^4 \times S^1$\cite{Benjamin:2017xen}. We continue the analysis of refined BPS spectra at the supergravity point in the moduli space of Sym$^N(M)$ at large $N$, analyzing the properties of the supergravity BPS spectrum upon decomposition into characters of the relevant worldsheet superconformal algebra of the dual CFT. The purpose of our paper is twofold:
\begin{enumerate}
\item  First we examine the large $N$ limit of 1/2-- and 1/4--BPS degeneracies of the HEG at the supergravity point; this applies to low--lying states of the dual CFT with conformal weight below the threshold corresponding to a black hole in the bulk. We show that in this limit the degeneracies ``stabilize"; i.e. they only depend on certain combinations of left-- and right--moving quantum numbers, which may be indicative of a new symmetry present.\footnote{Note that here we do not mean the usual definition of ``stabilization," whereby the low--lying spectrum is independent of $N$ for sufficiently large $N$. Instead we discuss a new phenomenon where the spectrum is invariant under the actions of (a) and (b) below.} We derive analytic expressions for the 1/2-- and 1/4--BPS degeneracies in this limit and find an unexpected appearance of Rogers Ramanujan functions in the 1/4--BPS case. Note that because quantum numbers of the right-movers are involved, such stabilization is not visible from just the EG;\footnote{In fact, the character decomposition of the elliptic genus does not even have a well--defined large $N$ limit.} only upon refinement does one observe this phenomenon. 

In particular, our results are as follows. Consider a 1/2--BPS state with left-- and right--moving NS-NS spins and conformal weights $(i,j)$, respectively. At large $N$ as long as $j$ is sufficiently large, the degeneracy of the representation is unchanged when we increase both the left-moving $SU(2)_R$ spin and conformal weight by $\frac12$, and increase both the right-moving $SU(2)_R$ spin and conformal weight by $\frac32$.
Furthermore, consider an NS-NS 1/4--BPS state with left-moving $SU(2)_R$ spin $i$ and conformal weight $\tilde h + i$; and both right-moving $SU(2)_R$ spin and conformal weight $j$ (so that the right-moving part is BPS). At large $N$ as long as $j$ is sufficiently large relative to $i$ and $\tilde h$, we find that the degeneracy of the representation is unchanged if we either:
\begin{enumerate}
\item Increase both the left-moving $SU(2)_R$ spin and conformal weight by $\frac12$, and increase both the right-moving $SU(2)_R$ spin and conformal weight by $\frac32$. (The same as observed for the 1/2--BPS degeneracies.)

\item Increase the left-moving conformal weight by $2$, and increase both the right-moving $SU(2)_R$ spin and conformal weight by $5$. (For the case of $M=T^4$, this symmetry is even larger: we can increase the left-moving conformal weight by $1$ and the right-moving $SU(2)_R$ spin and conformal weight by $\frac52$.)
\end{enumerate}
See equations (\ref{eq:stabcondk3}) and (\ref{eq:stabcondt4}) for the precise ranges of parameters in which these operations are symmetries.

\item Secondly, we consider the action of discrete symmetry groups on the refined supergravity BPS spectrum. In \cite{Benjamin:2016pil} it was observed that the coefficients in the character decomposition of the supergravity states counted by the HEG may have a connection to dimensions of irreducible representations of sporadic finite groups.\footnote{In particular, it was proposed that $M_{22}$ may be a symmetry.} This is reminiscent of the famous observation \cite{Eguchi:2010ej} that the character decomposition of the EG of a $K3$ surface can be decomposed into representations of the sporadic group $M_{24}$, a phenomenon now known as Mathieu moonshine. We show that there is indeed a connection between sporadic groups and the refined supergravity BPS spectrum, and it can be made precise for all four-plane preserving subgroups of the sporadic group $Co_0$. We explicitly define and compute the HEG of the bulk supergravity twined under  elements of such subgroups of $Co_0$, and discuss the possibility of a  larger discrete symmetry structure of the refined BPS spectrum at large $N$. We comment on a similar story for the case of $T^4$.

\end{enumerate}

The outline of the rest of the paper is as follows. In \S\ref{sec:spectrum} we review the spectrum of BPS states of both half--maximal and maximal supergravity on AdS$_3\times S^3$. In \S\ref{sec:theanswer} we explain point 1, and describe curious properties both of the 1/2--BPS spectra and of the supergravity 1/4--BPS spectra. In \S\ref{sec:symms}, we discuss point 2  in  detail. We present a general discussion of the action of discrete symmetry groups on states counted by the HEG  and derive explicit actions of four--plane--preserving subgroups of $Co_0$ on the refined 1/4--BPS spectrum of half--maximal supergravity. Finally in \S\ref{sec:discuss} we conclude and raise several potentially interesting questions. Some special functions and character formulae used in the text are given in Appendix \ref{app:thetachar}, and a few long derivations of the results in \S\ref{sec:theanswer} are presented in Appendix \ref{app:derivation}.

\section{The (refined) spectrum of supergravity on AdS$_3 \times S^3$}\label{sec:spectrum}
Here we briefly review the spectrum of half-maximal ($\mc N=(2,0)$) and maximal ($\mc N=(2,2)$) supergravity on AdS$_3 \times S^3$ (see \cite{deBoer:1998kjm, deBoer:1998us, Benjamin:2016pil} for more details). These supergravities arise upon compactification of type IIB string theory on AdS$_3\times S^3 \times M$, where $M=K3$ and $T^4$, respectively, and the radius of $M$ is much smaller than that of AdS$_3$ and $S^3$. These theories are holographically dual to a point in the moduli space of the CFT Sym$^N(M)$ as $N \to \infty$. 

Following \cite{deBoer:1998kjm, deBoer:1998us}, in \cite{Benjamin:2016pil} this relation was used to compute the low-lying spectrum of the HEG of Sym$^N(M)$ deformed to the supergravity point in the moduli space. In this section, we review the spectrum and decomposition of these supergravity states into characters of the relevant worldsheet superconformal algebra. We further study the symmetries of these degeneracies in \S \ref{sec:theanswer} and \S \ref{sec:symms}.

\subsection{Half--maximal supergravity}\label{sec:halfmaxspec}
The half--maximal $\mc N=(2,0)$ supergravity arises upon compactification of type IIB string theory on AdS$_3\times S^3 \times K3$, where the size of the $K3$ is much smaller than that of the AdS$_3$ and $S^3$. The dual 2d SCFT theory lies in the moduli space of Sym$^N(K3)$ and has $\mc N=(4,4)$ superconformal symmetry with central charge $c=6N$. This algebra has representations of two types: short (BPS) and long (non-BPS). We will specify the representations by the quantum numbers $h$ and $j$, the eigenvalues of the highest weight states under the operators $L_0$ and $J_0$ (the $SU(2)$ Cartan), respectively. In the NS sector there are $\frac c6+1$ familes of short representations which obey $h = \frac j2$, with $j \in \{0,1,\ldots,\frac c6\}$. We denote their characters by $\chi^{s, \text{NS}}_{j;\frac c6}$. Furthermore for any $h>\frac j2$, there are $\frac c6$ long representations with $j \in \{0,1,\ldots,\frac c6-1\}$. We first define $\tilde h\equiv h - \frac j2$ so that $\tilde h >0$ and denote their characters by $\chi^{\ell,\text{NS}}_{j,\tilde h; \frac c6}$. Similarly we denote the characters of the corresponding Ramond sector representations by  $\chi^{s, \text R}_{j;\frac c6}$ and $\chi^{\ell,\text{R}}_{j,\tilde h; \frac c6}$. These can be obtained from the NS characters via spectral flow. See Appendix \ref{app:smallchars} for explicit formulas for these characters.

We now turn to a discussion of the bulk spacetime. The Kaluza-Klein (KK) spectrum $\mathcal N=(2,0)$ supergravity on AdS$_3\times S^3$ can be organized into representations of the supergroup $SU(1,1|2)_L \times SU(1,1|2)_R$. This group is generated by the global part of the 2d $\mc N=(4,4)$ superconformal algebra. The KK spectrum consists of short representations of the supergroup which we denote by $(j,j')_S$. Such a representation corresponds to a chiral primary of the dual CFT in the NS-NS sector with $(J_0,\overline{J_0})$ eigenvalues $(j,j')$.

In \cite{deBoer:1998kjm} the KK spectrum of single--particle supergravity states was derived. Furthermore, in \cite{deBoer:1998us} it was understood that one can find a precise match at finite $N$ between degeneracies of 1/4--BPS states in the bulk supergravity and those counted by the EG of the dual CFT.  We review this relation now. First off, recall that the EG of a K3 non--linear sigma model (NLSM), which we refer to as $Z_{\rm EG}^{K3}$, has an expansion of the form
\be\label{eq:EGK3}
Z_{\text{EG}}^{K3}(\t,z)= \sum_{m,\ell} c(m,\ell) q^m y^\ell=\frac{2}y + 20 + 2 y + \mc O(q),
\ee
where the coefficients $c(m,\ell)\equiv c(4m-\ell^2)$ only depend on the combination of $4m -\ell^2$. In \cite{Dijkgraaf:1996xw} a generating function for the elliptic genera of the $N^{\text{th}}$ symmetric product of a K3 NLSM, Sym$^N(K3)$, was derived to be
\be\label{eq:DMVV}
\sum_{N=0}^\infty Z_{\rm EG}^{\text{Sym}^N(K3)}(\t,z) p^N = \prod_{n>0,m\geq 0, \ell} \frac{1}{(1-p^nq^my^\ell)^{c(mn,\ell)}} = \sum_{m,n,\ell}a(m,n,\ell) q^mp^n y^\ell,
\ee
where the $c(mn,\ell)$ are the coefficients in the expansion of the K3 EG (\ref{eq:EGK3}). In fact, as the EG is independent of the CFT moduli space, the formula (\ref{eq:DMVV}) holds for all points in the moduli space of the symmetric product theory.

From the supergravity side, one can reproduce the degeneracies $a(m,n,\ell)$ of the EG of the dual CFT for states with left--moving conformal weight $h \leq \frac{N+1}{4}$, i.e. below the threshold corresponding to black hole states \cite{deBoer:1998us}.\footnote{In fact, black hole states more precisely correspond to coefficients with ``polarity" $p\geq0$ where $p:=4mn -\ell^2$ \cite{Dijkgraaf:2000fq}. Thus, the supergravity analysis should apply in a slightly larger range of parameters than discussed in \cite{deBoer:1998us}.} This matching is achieved only after assigning a ``degree" to each short multiplet and imposing an exclusion principle following  \cite{Maldacena:1998bw} limiting the multi--particle states to have degree at most $N$. Labeling the short representations as $(j, j';d)$ where $d$ is the degree, the  spectrum of short multiplets including the degree is\cite{deBoer:1998us},
\begin{align}
&(n-1,n+1;n)_S   \nonumber\\
&(n+1, n+1;n)_S  \nonumber\\
&(n,n;n+1)_S  \nonumber\\
&(n+1,n-1;n)_S  \nonumber\\
&20(n,n;n)_S,
\label{eq:bearlasers}
\end{align}
where $n$ runs over positive integers. We denote the single--particle Hilbert space of degree $d$ as $\mathcal H^{(d),{\rm single}}_{(2,0)}$.

For a given $N$, the exclusion principle is implemented by considering the Hilbert space,
\be\label{eq:halfmaxHilb}
\mathcal H^{(N)}_{(2,0)}:= \bigoplus_{\substack{\{j_i,j_i';d_i\}\\ \sum d_i \leq N}}\bigotimes_i(j_i,j_i';d_i)_S=\bigoplus_{\substack{\{j_i,j_i';d_i\}\\ \sum d_i \leq N}}\bigotimes_i \mathcal H^{(d_i),{\rm single}}_{(2,0)}.
\ee
The supergravity EG is given by the trace,
\be
\widetilde Z^{(2,0)}_{\text{EG},N}(\t,z):=\Tr_{\mathcal H^{(N)}_{(2,0)}} \((-1)^{F} q^{L_0} y^{J_0}\)\Big |_{\rm right-chiral},
\ee
where we trace only over states that are chiral primaries on the right.  We use a tilde to denote the fact that we are doing a supergravity computation. An explicit generating function for $\widetilde Z^{(2,0)}_{\text{EG},N}(\t,z)$ can be found in \cite{deBoer:1998us}. One finds a matching of degeneracies in the relevant range of parameters after implementing a spectral flow of equation (\ref{eq:DMVV}) to the NS-NS sector and comparing the coefficients of the  power series expansion of the two generating functions. We refer to \cite{deBoer:1998us} for details.

In \cite{Benjamin:2016pil} a similar analysis was performed for the supergravity spectrum including a refinement by the right--moving $U(1)$ current. In this case, one cannot compare degeneracies with the HEG of the dual CFT as it is not known independently. However, by analogous reasoning the trace\footnote{Our convention is that in (\ref{eq:trwhatever}), $\overline{J_0}$ is taken in the R sector which differs by $c/6$ from the charge in the NS sector.}
\be
\widetilde Z^{(2,0)}_{\text{HEG},N}(\t,z,\nu):=\Tr_{\mathcal H^{(N)}_{(2,0)}} \((-1)^{F} q^{L_0} y^{J_0} u^{\overline{J_0}}\)\Big |_{\rm right-chiral},
\label{eq:trwhatever}
\ee
(where again we only trace over right-moving chiral primaries) captures degeneracies of the HEG of the dual CFT at finite $N$ for conformal primaries with eigenvalues below the threshold corresponding to black hole states. An explicit generating function for such states was derived in \cite{Benjamin:2016pil}. The result is given by
\be
\sum_{n\geq0}\widetilde Z^{(2,0)}_{\text{HEG},N}(\t,z,\nu) p^n = \prod_{n>0,m, \ell, \ell'}\frac1{(1-p^nq^my^\ell u^{\ell'})^{c^{(2,0)}_{\text{sugra}}(n,m,\ell,\ell')}},
\label{eq:fullheg}
\ee
where the coefficients $c^{(2,0)}_{\text{sugra}}$ are defined by
\begin{align}
\sum_{N=1}^\infty &\left (\Tr_{\mathcal H^{(N),{\rm single}}_{(2,0)}}(-1)^F q^{L_0}y^{J_0}u^{\overline J_0}\right )\Bigg|_{\rm right-chiral}p^N =\sum_{n,m,\ell,\ell'} c^{(2,0)}_{\text{sugra}}(n,m,\ell,\ell') p^n q^m y^\ell u^{\ell'}  \nn\\
&=\frac{1}{(1-q)(y-y^{-1})} \bigg(\frac{(u+u^{-1})p^2}{1-q^{1/2}yp}(y^2 q^{1/2} - 2 yq + q^{3/2}) \nn\\
&~~-\frac{(u+u^{-1})p^{2}}{1-q^{1/2}y^{-1}p}(y^{-2}q^{1/2}-2y^{-1}q+q^{3/2})+\frac{(u+u^{-1})p}{1-q^{1/2}yp}(y^3q-2y^2q^{3/2}+yq^2)\nonumber\\
&~~-\frac{(u+u^{-1})p}{1-q^{1/2}y^{-1}p}(y^{-3}q-2y^{-2}q^{3/2}+y^{-1}q^2)+\frac{20p}{1-q^{1/2}yp}(y^2q^{1/2}-2yq+q^{3/2})\nn\\
&~~-\frac{20p}{1-q^{1/2}y^{-1}p}(y^{-2}q^{1/2} -2y^{-1}q+q^{3/2})\bigg)+(u+u^{-1})p \nn\\
&\equiv f^{(2,0)}(p, q, y, u).
\label{eq:sugra}
\end{align}
Note that the generating function (\ref{eq:fullheg}) corresponds to states in the NS-R sector of the dual CFT \cite{Benjamin:2016pil}.

Given that the dual CFT has $\mc{N}=(4,4)$ supersymmetry, it follows that $\widetilde Z^{(2,0)}_{\text{HEG},N}$ has a character decomposition of the form
\be
\widetilde Z^{(2,0)}_{\text{HEG},N}(\t,z,\nu) = \sum_{j,\overline j=0}^N c_N^{j,\overline j} \chi^{s,\text{NS}}_{j; N}(\t,z) \overline{\chi^{s,\text{R}}_{\overline j;N}}(\overline \t,\nu)+\sum_{j=0}^{N-1}\sum_{\overline j = 0}^{N}\sum_{\tilde h = 1}^\infty c_N^{j,\tilde h,\overline j} \chi^{\ell,\text{NS}}_{j,\tilde h; N}(\t,z) \overline{\chi^{s,\text{R}}_{\overline j;N}}(\overline \t,\nu)\label{eq:decompk3}
\ee
where $c_N^{j,\overline j}$ and $c_N^{j,\tilde h,\overline j}$ denote the degeneracies of 1/2-- and 1/4--BPS states, respectively. 
At large $N$ the generating function (\ref{eq:fullheg}) and character formulas are independent of $N$, and the decomposition ``stabilizes." The first few such coefficients in this expansion are \cite{Benjamin:2016pil}:
\begin{align}
\widetilde Z^{(2,0)}_{\text{HEG}}(\tau, z, \nu) &\stackrel{{\rm large~} N}{\sim} \( 210 \overline{\chi^{s}_2} + 21\overline{\chi^{s}_4}\) \chi^{\ell}_{0,1} + \(3542 \overline{\chi^s_3} + 484 \overline{\chi^s_5} + 22 \overline{\chi^s_7} \) \chi^{\ell}_{1,1} \nn\\&~~+ \( 21 \overline{\chi_2^s} + 36961\overline{\chi_4^s} + 6281\overline{\chi_6^s} + 506 \overline{\chi_8^s} + 22\overline{\chi_{10}^s} \) \chi^{\ell}_{2,1} \nn\\& ~~ + \( 231 \overline{\chi_2^s} + 2660 \overline{\chi_3^s} + 21526\overline{\chi_4^s} + 420 \overline{\chi_5^s} + 3796 \overline{\chi_6^s} + 275 \overline{\chi_8^s} + \overline{\chi_{10}^s}\) \chi^{\ell}_{0,2} \nn\\ &~~+ \ldots.
\label{eq:sushi}
\end{align}
In (\ref{eq:sushi}) we include only the 1/4--BPS spectrum (not the 1/2--BPS spectrum) up to $\mc{O}(q^2)$ above the vacuum. For clarity we drop the subscript of $N$ to indicate we have taken the large $N$ limit and suppress the superscripts of NS and R. In \S \ref{sec:theanswer} we discuss a further stablization property in more detail.

\subsection{Maximal supergravity}\label{sec:maxspec}
In this section we discuss the case of $\mathcal N=(2,2)$ supergravity which arises after type IIB compactification on AdS$_3 \times S^3 \times T^4$. The dual SCFT lies in the moduli space of $\text{Sym}^N(T^4)$,  has central charge $c=6N$, and $\mathcal N=(4,4)$ superconformal symmetry, where the $\mathcal N=4$ is now a Wigner contraction of the large $\mathcal N=4$ superconformal algebra \cite{Sevrin:1988ew, Maldacena:1999bp}. As in the case of the small $\mathcal N=4$ algebra, we can specify representations by $h$ and $j$, the eigenvalues of $L_0$ and $J_0$, the $SU(2)$ Cartan. There are short representations which have $h= \frac{j}2$, $j \in \{0,\ldots \frac{c}6-1\}$ whose characters we will denote by $\check \chi^{s,\text{NS}}_{j;\frac{c}6}$ and $\check \chi^{s,\text{R}}_{j;\frac{c}6}$. The long representations have $h>\frac{j}2$ and $j \in \{0,\ldots \frac{c}6-2\}$; defining again $\tilde h \equiv h-\frac{j}2$, we denote their characters by $\check \chi^{\ell,\text{NS}}_{j,\tilde h;\frac{c}6}$ and $\check \chi^{\ell,\text{R}}_{j,\tilde h;\frac{c}6}$. Explicit formulas for these characters are given in Appendix \ref{app:largen4}.

The KK spectrum of $\mathcal N=(2,2)$ supergravity on AdS$_3\times S^3$ was discussed in \cite{deBoer:1998kjm}. It can again be organized into representations of $SU(1,1|2)_L \times SU(1,1|2)_R$; however, there are now multiplets with fermionic highest weight states. Using the same labeling as in the previous section, the single--particle spectrum including the degree is then given by the data in Table \ref{tbl:maxmults}, where  $n$ runs over positive integers.

\begin{table}[htb]
\begin{center}
\begin{tabular}{c|c}
Bosonic multiplets & Fermionic multiplets \\\hline
$\begin{aligned}
&(n, n; n+1)_S \nn\\
&(n-1,n+1;n)_S \nn\\
&(n+1,n-1;n)_S\nn\\
&(n+1,n+1;n)_S\nn\\
&4(n,n;n)_S
\label{eq:caterpillarlaser}
\end{aligned}$
&
$\begin{aligned}
&2(n-1, n;n)_S \nn\\
&2(n,n-1;n)_S \nn\\
&2(n,n+1;n)_S\nn\\
&2(n+1,n;n)_S
\label{eq:caterpillarlaser2},
\end{aligned}$
\end{tabular}
\caption{Spectrum of single--particle $\mathcal N=(2,2)$ supergravity multiplets.}\label{tbl:maxmults}
\end{center}
\end{table}

The relation between the supergravity spectrum and that of the dual CFT is similar to the discussion in the previous section, so we will be brief. We again denote the single--particle Hilbert space by $\mathcal H^{(d),{\rm single}}_{(2,2)}$. The relevant supergravity Hilbert space at finite $N$ including multi--particle states is given by
\be\label{eq:maxHilb}
\mathcal H^{(N)}_{(2,2)}:= \bigoplus_{\substack{\{j_i,j_i';d_i\}\\ \sum d_i \leq N}}\bigotimes_i(j_i,j_i';d_i)_S=\bigoplus_{\substack{\{j_i,j_i';d_i\}\\ \sum d_i \leq N}}\bigotimes_i \mathcal H^{(d_i),{\rm single}}_{(2,2)},
\ee
where we have again implemented an exclusion principle limiting the spectrum at finite $N$ to include (single-- and multi--particle) states with degree at most $N$. The EG of the dual CFT vanishes due to fermionic zero modes; however, one can see matching of the supergravity BPS spectrum with the spectrum of the dual CFT at finite $N$ by considering a modified index \cite{Maldacena:1999bp}.

On the other hand, the refined supergravity EG does not vanish and can be organized into a generating function given by
\be
\sum_{n\geq0} \widetilde Z_{\text{HEG},N}^{(2,2)}(\tau,z,\nu) p^n = \prod_{n>0,m, \ell, \ell'}\frac1{(1-p^nq^my^\ell u^{\ell'})^{c_{\text{sugra}}^{(2,2)}(n,m,\ell,\ell')}},
\label{eq:fullhegT4}
\ee
where the coefficients $c_{\text{sugra}}^{(2,2)}$ are defined via
\begin{align}
\sum_{N=1}^\infty &\left (\Tr_{\mathcal H^{(N),{\rm single}}_{(2,2)}}(-1)^F q^{L_0}y^{J_0}u^{\overline J_0}\right )\Bigg|_{\rm right-chiral}p^N=\sum_{n,m,\ell,\ell'} c_{\text{sugra}}^{(2,2)}(n,m,\ell,\ell') p^n q^m y^\ell u^{\ell'}= \nonumber\\ 
&\frac{1}{(1-q)(y-y^{-1})} \bigg(\frac{(u+u^{-1}-2)p^2+(4-2u^{-1}-2u)p}{1-q^{1/2}yp}(y^2 q^{1/2} - 2 yq + q^{3/2}) \nn\\
&~~-\frac{(u+u^{-1}-2)p^2 + (4-2u^{-1}-2u)p}{1-q^{1/2}y^{-1}p}(y^{-2}q^{1/2}-2y^{-1}q+q^{3/2})\nn\\&~~+\frac{(u+u^{-1}-2)p}{1-q^{1/2} yp}(y^3 q-2y^2 q^{3/2}+y q^2)-\frac{(u+u^{-1}-2)p}{1-q^{1/2}y^{-1}p}(y^{-3}q-2y^{-2}q^{3/2}+y^{-1}q^2)\bigg)\nn\\&~~+(u+u^{-1}-2)p \nn\\
&\equiv f^{(2,2)}(p,q,y,u),
\label{eq:t4sugrakkearly}
\end{align} 
and, again, the states in equation (\ref{eq:fullhegT4}) are understood to correspond to the NS-R sector of the dual CFT \cite{Benjamin:2016pil}.

Given the superconformal symmetry of the dual CFT, it follows that $\widetilde Z_{\text{HEG},N}^{(2,2)}$ has a decomposition of the form
\be
\widetilde Z^{(2,2)}_{\text{HEG},N}(\t,z,\nu) = \sum_{j,\overline j=0}^{N-1} \check c_N^{j,\overline j} \check \chi^{s,\text{NS}}_{j; N}(\t,z) \overline{\check \chi^{s,\text{R}}_{\overline j;N}}(\overline \t,\nu)+\sum_{j=0}^{N-2}\sum_{\overline j = 0}^{N-1}\sum_{\tilde h = 1}^\infty \check c_N^{j,\tilde h,\overline j} \check \chi^{\ell,\text{NS}}_{j,\tilde h; N}(\t,z) \overline{\check\chi^{s,\text{R}}_{\overline j;N}}(\overline \t,\nu)\label{eq:decompt4}
\ee
into Wigner--contracted large $\mathcal N=4$ superconformal characters. 
Explicitly, the first few coefficients in this decomposition at large $N$ are given by
\begin{align}
\widetilde Z_{\text{HEG}}^{(2,2)}&(\tau, z, \nu) \stackrel{{\rm large} ~N}{\sim} \(  10\overline{\check \chi^{s}_2} + 20\overline{\check\chi^{s}_3} + 15\overline{\check\chi^{s}_4} + 4\overline{\check\chi^{s}_5}\) \check\chi^{\ell}_{0,1} \nn\\&
~~+ \( 20\overline{\check\chi^s_2} + 86 \overline{\check\chi^s_3} +  148\overline{\check\chi^s_4}+  132\overline{\check\chi^s_5}+ 68\overline{\check\chi^s_6}+  22\overline{\check\chi^s_7}+  4\overline{\check\chi^s_8} \) \check\chi^{\ell}_{1,1}\nn\\&
~~+ \( 15\overline{\check\chi^s_2} + 148 \overline{\check\chi^s_3} +  493\overline{\check\chi^s_4}+  820\overline{\check\chi^s_5}+ 791\overline{\check\chi^s_6}+  488\overline{\check\chi^s_7}+  215\overline{\check\chi^s_8} +  76\overline{\check\chi^s_9} +  22\overline{\check\chi^s_{10}} +  4\overline{\check\chi^s_{11}} \) \check\chi^{\ell}_{2,1} \nn\\&
~~ + \(  21\overline{\check\chi_2^s} +  124\overline{\check\chi_3^s} + 348\overline{\check\chi_4^s} +  536\overline{\check\chi_5^s} +  500\overline{\check\chi_6^s} +  316\overline{\check\chi_7^s} + 149\overline{\check\chi_8^s} + 48\overline{\check\chi_{9}^s} + 7\overline{\check\chi_{10}^s}\) \check\chi^{\ell}_{0,2} \nn\\ &~~+ \ldots.
\label{eq:sashimi}
\end{align}
In (\ref{eq:sashimi}) we include only the 1/4--BPS states up to $\mc{O}(q^2)$ above the vacuum. We again suppress the subscript $N$ to indicate we have taken the large $N$ limit and the superscripts NS, R for clarity.

\section{Stabilization of degeneracies at large $N$} 
\label{sec:theanswer}

In this section we point out some observations about the degeneracies of (supergravity) 1/2-- and 1/4--BPS states in the large $N$ limit. For CFTs in the moduli space of Sym$^N(M)$, denote by $c_N^{i,j}, \check c_N^{i,j}$ the degeneracies of 1/2--BPS states with $(J_0, \overline{J_0})$ eigenvalues $(i,j)$ for $M=K3$ and $T^4$, respectively. We find that after taking $N\to \infty$, the degeneracies $c^{i,j}_N, \check c^{i,j}_N$ only depend on the combination $3i-j$ of the left-- and right--moving $U(1)$ charges when $j$ is sufficiently large.

Similarly, denote by $c_N^{i,\tilde h;j}$ and $\check c_N^{i,\tilde h;j}$ the degeneracies of 1/4--BPS states with $(L_0, \overline{L_0}, J_0, \overline{J_0})$ eigenvalues $(\frac i2+\tilde h, \frac j2, i,j)$ in the NS-NS sector for $M=K3$ and $T^4$ respectively. We find that after first taking the large $N$ limit, the degeneracies $c_N^{i,\tilde h;j}$ only depend on the combination $3i + 5\tilde h-j$ and the parity of $\tilde h$, for sufficiently large $j$. Similarly the degeneracies $\check c_N^{i, \tilde h; j}$ depend only on the combination $3i+5\tilde h -j$. We refer to the above phenomena as ``stabilization." Furthermore, we  derive explicit formulas for these degeneracies in the range where they have stabilized. Note that this phenomenon only occurs upon considering the fully refined 1/4--BPS spectrum; the degeneracy of states counted by the EG never stabilizes but rather grows linearly with $N$ due to right-moving ground state degeneracies \cite{deBoer:1998kjm}. In this section we simply state the results; for the derivations see Appendix \ref{app:derivation}.

\subsection{Half--BPS degeneracies}\label{sec:halfstab}
First we discuss the character decomposition of the refined spectrum of 1/2--BPS states at large $N$. As the spectrum of 1/2--BPS states is moduli--independent, the results of this section apply at any point in the moduli space of Sym$^N(M)$. For $M=K3$ the generating function of 1/2--BPS states (in the RR sector) refined by both left-- and right--moving $U(1)$ charges can be obtained, e.g., from a formula in \cite{Gottsche}: 
\be\label{eq:hodgeK3}
\sum_{N=0}^{\infty} p^N Z_{\text{Hodge}}^{\text{Sym}^N(K3)}(z,\nu) = \prod_{n=1}^{\infty} \frac{1}{(1-p^n)^{20}(1-p^nuy)(1-p^nu^{-1}y)(1-p^nuy^{-1})(1-p^nu^{-1}y^{-1})},
\ee
where for $M$ a $d$-dimensional K\"ahler manifold, $Z_{\text{Hodge}}^{\text{Sym}^N(K3)}(z,\nu)$ denotes its Hodge polynomial, i.e.
\be
Z_{\text{Hodge}}^{M}(z,\nu):=u^{-d/2}y^{-d/2} \sum_{p,q} (-u)^q(-y)^p h^{p,q}(M).
\ee

We consider the decomposition of $Z_{\text{Hodge}}$ into $\mc{N}=4$ characters. Because the representations  are RR ground states, the only characters that can contribute are those of the form (short, short). Thus we can write
\be
Z_{\text{Hodge}}^{\text{Sym}^N(K3)}(z,\nu) = \sum_{i,j=0}^N c^{i,j}_N \chi^{s,\text R}_{i;N}(z) \overline{\chi^{s,\text R}_{j;N}}(\nu),
\ee
where the coefficient $c^{i,j}_N$ denotes the number of 1/2--BPS mutliplets with $(J_0, \overline{J_0})$ eigenvalues $(i,j)$ for all CFTs in the moduli space of Sym$^N(K3)$.

In a certain regime, these degeneracies are independent of $N$ and depend on a single combination of left-- and right--moving spin. Let $c^{i,j} := \lim_{N\to\infty} c_N^{i,j}.$\footnote{This limit was shown to be well-defined in \cite{deBoer:1998kjm}.} Define $k:= \frac 32 i -\frac 12j$.
In Appendix \ref{sec:derivK3half} we show that the degeneracies $c^{i, 3i-2k}$ only depend on $k$ after first taking the limit $N \to\infty$ and then the limit $i\to\infty$. Moreover  they are given by the generating function
\begin{align}\label{eq:K3stab1}
\sum_{k=0}^{\infty} c^{i, 3i-2k} q^k &= (1-q)^2(1-q^2)(1-q^3)\prod_{n=1}^{\infty}\frac1{(1-q^n)^{24}}\nn\\ 
&=q(1-q)^2(1-q^2)(1-q^3)\frac{1}{\eta(\t)^{24}} \nn \\
&= 1 + 22q + 276q^2 + 2553q^3 + 19275q^4 + 125304q^5 + \mc{O}(q^6).
\end{align}
In fact, we find empirically that this generating function captures the degeneracies $c_N^{i, 3i-2k}$ at finite $N$ as long as 
$N\geq 3i$ and $i\geq 2k$ (see Table \ref{tab:halfbpsk3}).

\begin{table}[h]
\begin{center}
\def\arraystretch{1.2}
\begin{tabular}{| c || c | c | c | c | c | c |}
\hline
& $\chi_{i}^s \overline{\chi_{3i}^s}$& $\chi_{i}^s\overline{ \chi_{3i-2}^s}$ & $\chi_{i}^s \overline{\chi_{3i-4}^s}$ & $\chi_{i}^s \overline{\chi_{3i-6}^s}$ & $\chi_{i}^s\overline{\chi_{3i-8}^s}$ & $\chi_{i}^s \overline{\chi_{3i-10}^s}$\\
\hline
$i=0$ & {\bf 1} & -- & -- & -- &  -- & -- \\
\hline
$i=1$ & {\bf 1}  & 21 & --  & -- & --& --\\
\hline
$i=2$ & {\bf 1}  & {\bf 22} & 253  & 0 & -- & -- \\
\hline
$i=3$ & {\bf 1}  & {\bf 22} & 275  & 2255 & 1  & --  \\
\hline
$i=4$ & {\bf 1}  &{\bf 22} & {\bf 276}  & 2530  &  16446 &22   \\
\hline
$i=5$ & {\bf 1}  & {\bf 22} & {\bf 276}  & 2552  & 18976  & 103478 \\
\hline
$i=6$ & {\bf 1}  & {\bf 22} & {\bf 276}  & {\bf 2553}  & 19252  & 122453 \\
\hline
$i=7$ & {\bf 1}  & {\bf 22} & {\bf 276}  & {\bf 2553}  & 19274  & 125005 \\
\hline
$i=8$ & {\bf 1}  & {\bf 22} & {\bf 276}  & {\bf 2553}  & {\bf 19275}  & 125281 \\
\hline
$i=9$ & {\bf 1}  & {\bf 22} & {\bf 276}  & {\bf 2553}  & {\bf 19275}  & 125303 \\
\hline
$i=10$ & {\bf 1}  & {\bf 22} & {\bf 276}  & {\bf 2553}  & {\bf 19275}  & {\bf 125304} \\
\hline
\end{tabular}
\end{center}
\caption{Degeneracies of 1/2--BPS states of Sym$^N(K3)$ at large $N$. Bold numbers indicate the degeneracy has stabilized and is computed by the coefficients of (\ref{eq:K3stab1}).}
\label{tab:halfbpsk3}
\end{table}

Similarly, we study the large $N$ limit of the degeneracies of 1/2--BPS states on Sym$^N(T^4)$ and its resolutions. The generating function for these states has the form\cite{Gottsche} 
\be
\sum_{N=0}^{\infty} p^N Z_{\text{Hodge}}^{\text{Sym}^N(T^4)}(z,\nu)  = \prod_{n=1}^{\infty} \frac{(1-p^nu^{-1})^2(1-p^nu)^2(1-p^ny^{-1})^2(1-p^ny)^2}{(1-p^n)^4(1-p^nu^{-1}y^{-1})(1-p^nu^{-1}y)(1-p^nuy^{-1})(1-p^nuy)}.
\label{eq:hodgeT4}
\ee
We decompose the Hodge polynomial of Sym$^N(T^4)$ into contracted large $\mc N=4$ characters in the following way:
\be
Z_{\text{Hodge}}^{\text{Sym}^N(T^4)}(z,\nu) = \sum_{i,j=0}^{N-1} \check c^{i,j}_N \check \chi^{s,R}_{i;N}(z) \overline{\check \chi^{s,R}_{j;N}}(\nu),
\ee
where now $\check c^{i,j}_N$ denotes the degeneracy of 1/2--BPS mutliplets with $(J_0, \overline{J_0})$ eigenvalues $(i,j)$ for all CFTs in the moduli space of Sym$^N(T^4)$.

Defining $\check c^{i,j}:= \lim_{N\to \infty} \check c^{i,j}_N$, we again find that in a certain regime, $\check c^{i,3i-2k}$ only depends on the combination $k$ as defined above. Furthermore, in Appendix \ref{sec:derivT4half} we derive an explicit generating function for the degeneracies in this limit given by
\begin{align}\label{eq:T4stab1}
\sum_{k \el \frac12 \bb Z} \check c^{i, 3i-2k}_n q^{k} 
&= \frac{(1-q)^2(1-q^2)(1-q^3)}{(1+q^{\frac12})^4(1+q^{\frac32})^4}\prod_{n=1}^{\infty} \frac{(1+q^{n-\frac12})^8}{(1-q^n)^8} 
\nn\\ 
&= \frac{q^{\frac12}(1-q)^2(1-q^2)(1-q^3)}{(1+q^{\frac12})^4(1+q^{\frac32})^4}\frac{\eta(\t)^8}{\eta\left (\frac{\tau}2\right )^8 \eta(2\t)^8}
\nn\\
&= 1 + 4q^{\frac12} + 12q + 32q^{\frac32} + 81q^2 + 192q^{\frac52} + 429q^3 + 920q^{\frac72} + \mc{O}(q^4).
\end{align}
Again we find empirically that this generating function describes the degeneracies $\check c_N^{i, 3i-2k}$ at finite $N$ whenever $N\geq 3i+1$ and $i\geq 2k$ (see Table \ref{tab:halfbpst4}).

\begin{table}[h]
\begin{center}
\def\arraystretch{1.2}
\begin{tabular}{| c || c | c | c | c | c | c |}
\hline

&  $\check\chi_{i}^s \overline{\check \chi_{3i}^s}$ & $\check\chi_{i}^s\overline{\check \chi_{3i-1}^s}$ & $\check\chi_{i}^s \overline{\check\chi_{3i-2}^s}$ & $\check\chi_{i}^s \overline{\check\chi_{3i-3}^s}$ & $\check\chi_{i}^s\overline{\check\chi_{3i-4}^s}$ & $\check\chi_{i}^s \overline{\check\chi_{3i-5}^s}$\\
\hline
$i=0$ & {\bf 1} & -- & -- & -- &  -- & -- \\
\hline
$i=1$ & {\bf 1}  & {\bf 4} & 5  & 0 & --& --\\
\hline
$i=2$ & {\bf 1}  & {\bf 4} & {\bf 12}  & 24 & 21 & 4 \\
\hline
$i=3$ & {\bf 1}  & {\bf 4} & {\bf 12}  & {\bf 32} & 73  & 112  \\
\hline
$i=4$ & {\bf 1}  &{\bf 4} & {\bf 12}  & {\bf 32}  &  {\bf 81} &184   \\
\hline
$i=5$ & {\bf 1}  & {\bf 4} & {\bf 12}  & {\bf 32}  & {\bf 81}  & {\bf 192} \\
\hline
\end{tabular}
\end{center}
\caption{Degeneracies of half-BPS states of Sym$^N(T^4)$ at large $N$. Bold numbers indicate the degeneracy has  stabilized and is computed by the coefficients of (\ref{eq:T4stab1}).}
\label{tab:halfbpst4}
\end{table}

\subsection{Quarter--BPS degeneracies} \label{sec:quartstab}

In this section we summarize the results of a similar analysis for 1/4--BPS states of the large $N$ limit of Sym$^N(M)$ deformed to the supergravity point in moduli space. As the refined spectrum of 1/4--BPS states is moduli-dependent, the results of this section only apply to the supergravity locus in moduli space. As in equations (\ref{eq:decompk3}) and (\ref{eq:decompt4}), we denote by $c_N^{i, \tilde h, j},\check c_N^{i, \tilde h, j}$ the degeneracy of representations of the form
\be
\chi^{\ell}_{i, \tilde h;N} \overline{\chi^{s}_{j;N}}; ~~~\check \chi^{\ell}_{i, \tilde h;N} \overline{\check \chi^{s}_{j;N}}
\ee
for $M=K3$ and $M=T^4$, respectively. Furthermore we find that, after taking the limit $N\to \infty$, the degeneracies are independent of $N$, and at large $j$, they only depend on a linear combination of $i, \tilde h, j$ (as well as the parity of $\tilde h$ for the case of $M=K3$). In particular, they depend only on the linear combination $3i + 5\tilde h - j$, and, for $K3$, whether $\tilde h$ is even or odd (see (\ref{eq:stabcondk3}) and (\ref{eq:stabcondt4}) for the precise condition on the parameters necessary).

Note that this means that in, say, the NS-NS sector, if we (a) increase the left-moving spin and conformal weight by $\frac12$ and the right-moving spin and conformal weight by $\frac32$; or (b) increase the left-moving conformal weight by $2$ and the right-moving spin and conformal weight by $5$, the degeneracy remains the same, assuming $j$ is large enough.\footnote{In our notation, these operations correspond to (a) increasing $i,j$ by $1,3$, resp.; and (b) $\tilde h,j$ by $2,10$, resp. These are in some sense the ``minimal" operations which leave the degeneracies invariant. For the case of $T^4$, as the parity of $\tilde h$ does not play a role, the minimal version of (b) is instead increasing $\tilde h,j$ by $1,5$, resp.} Note that (a) is the same symmetry that we found for the 1/2--BPS degeneracies in the previous section.

In particular let $k':=3i+5\tilde h-j$, $c^{i, \tilde h, j}:=\lim_{N\to \infty}c_N^{i, \tilde h, j}$,  $\check c^{i, \tilde h, j}:=\lim_{N\to \infty}\check c_N^{i, \tilde h, j}$. Then, explicitly, we find the following set of generating functions for the degeneracies in this limit. For the case of $K3$, we derive a formula for the difference of even and odd $\tilde h$ degeneracies, given by,
\begin{align}
\sum_{k'} &\Big | c^{i, \tilde h, 3i+5\tilde h-k'} - c^{i, \tilde h+1, 3i+5(\tilde h+1)-k'}\Big | q^{k'} \nn\\ &= (1-q)^2(1-q^2)\prod_{n=1}^{\infty}\frac{1}{(1-q^{5n-4})^{25}(1-q^{5n-3})^{47}(1-q^{5n-2})^{47}(1-q^{5n-1})^{25}(1-q^{5n})^{24}}\nn\\
&= \frac{q^{\frac65}(1-q)^2(1-q^2)}{R(\tau)} \(\frac{H(\tau)}{\eta(\tau)}\)^{24} \nn\\
&= 1 + 23q + 322q^2 + 3405 q^3 + 29833q^4 + 227126q^5 +  \mc{O}(q^6)
\label{eq:ansdiffk3}
\end{align}
as well as a formula for the sum of even and odd $\tilde h$ which is is given by 
\begin{align}
\sum_{k'} & \(c^{i, \tilde h, 3i+5\tilde h-k'} + c^{i, \tilde h+1, 3i+5(\tilde h+1)-k'}\) q^{k'} \nn\\ &= \psi(\t)\left (\beta(\t)H(\t)\frac{R(2\t)^{\frac25}}{R(\t)^{\frac45}}\frac{\eta(4\t)^{\frac12}}{\eta(2\t)^{\frac32}}\right )^{24}\nn\\
&= 1 + 23q + 322q^2 + 3405 q^3 + 29925q^4 + 229338q^5 + \mc{O}(q^6),
\label{eq:ansansdiffsk3}
\end{align}
where we make use of the following definitions,
\begin{align}
G(\t) &:= \prod_{n=1}^{\infty} \frac{1}{(1-q^{5n-4})(1-q^{5n-1})} \nn\\
H(\t) &:= \prod_{n=1}^{\infty} \frac{1}{(1-q^{5n-3})(1-q^{5n-2})} \nn\\
R(\t) &:= q^{\frac15} \frac{H(\t)}{G(\t)} \nn\\
\psi(\t)&:=q(1-q)(1+q)^3 \frac{G(2\t)^2}{G(\t)^3H(\t)} \nn\\
\beta(\t)&:=\prod_{n=1}^{\infty} \frac{(1+q^n)^{\frac{n^2}{10}}}{(1-q^n)^{\frac{n^2}{10}}}.
\label{eq:psidef}
\end{align}
The functions $G(\tau)$ and $H(\tau)$ are called Rogers-Ramanujan functions; for more details see Appendix \ref{app:thetachar}. 

For the case of $T^4$ the generating function is independent of the parity of $\tilde h$ and is given by
\begin{align}
\sum_{k'} &\check c^{i, \tilde h, 3i+5\tilde h-k'} q^{k'} \nn\\ &= \widetilde \psi(\tau) \( \beta(\t) H(\t) \frac{R(2\t)^{\frac25}}{R(\t)^{\frac45}}\frac{\eta(2\tau)^{\frac12}}{\eta(\t)^{\frac12}}\)^{16}\nn\\
&= 8 + 120q + 1072q^2 + 7432 q^3 + 43896q^4 + 231136q^5 + \mc{O}(q^6).
\label{eq:t4sumwoo}
\end{align}
where we define,
\be
\widetilde \psi(\tau) := \frac{8(1-q)}{q^{\frac13}(1+q)}\frac{G(2\tau)^6}{G(\tau)^7H(\tau)}.
\ee

The above equations (\ref{eq:ansdiffk3}) and (\ref{eq:ansansdiffsk3}) are valid as long as
\begin{align}
i &\geq k' \nn \\ 
\tilde h &\geq 3k'-1. 
\label{eq:stabcondk3}
\end{align}
and equation (\ref{eq:t4sumwoo}) holds as long as 
\begin{align}
i &\geq k' \nn \\ 
\tilde h &\geq 3k'+3.
\label{eq:stabcondt4} 
\end{align}
Detailed derivations can be found in Appendix \ref{app:derivation}.\footnote{In our derivations, we take the limit as $i, \tilde h \rightarrow\infty$, with $i\gg\tilde h$, and held $k'$ constant. However from empirical data, we believe (\ref{eq:ansdiffk3}) and (\ref{eq:ansansdiffsk3}) should hold anytime (\ref{eq:stabcondk3}) is satisifed; similarly  (\ref{eq:t4sumwoo}) should hold anytime (\ref{eq:stabcondt4}) is satisfied.}

As an example, in Table \ref{tab:regulartable} we write down some terms in the character decomposition of Sym$^N(K3)$ deformed to the supergravity point. We can see that as we increase $\tilde h$, eventually the rows ``stabilize" depending on the parity of $\tilde h$.  Similarly, in Table \ref{tab:regulartableT4} we present an analogous table for $T^4$; note that in Table \ref{tab:regulartableT4} there is no dependence on the parity of $\tilde h$.

\begin{table}[h]
\begin{center}
\def\arraystretch{1.2}
\begin{tabular}{| c || c | c | c | c | c | c |}
\hline

&  $\chi_{i,\tilde h}^\ell \overline{\chi_{3i+5\tilde h}^s}$ & $\chi_{i,\tilde h}^\ell\overline{ \chi_{3i+5\tilde h-1}^s}$ & $\chi_{i,\tilde h}^\ell \overline{\chi_{3i+5\tilde h-2}^s}$ & $\chi_{i,\tilde h}^\ell \overline{\chi_{3i+5\tilde h-3}^s}$ & $\chi_{i,\tilde h}^\ell\overline{\chi_{3i+5\tilde h-4}^s}$ & $\chi_{i,\tilde h}^\ell \overline{\chi_{3i+5\tilde h-5}^s}$\\
\hline
$\frac12$-BPS, $\tilde h=0$ & {\bf 1} & 0 & 22 & 0 &  276 & 0 \\
\hline
$\frac14$-BPS, $\tilde h=1$ & 0  & 22 & 0  & 507 & 0 & 6601\\
\hline
$\frac14$-BPS, $\tilde h=2$ & {\bf 1}  & 0 & 298  & 0 & 6878 & 462 \\
\hline
$\frac14$-BPS, $\tilde h=3$ & 0  & {\bf 23} & 0  & 3082 & 44 & 69944\\
\hline
$\frac14$-BPS, $\tilde h=4$ & {\bf 1}  &0 & 321  & 0 & 26681 & 1058 \\
\hline
$\frac14$-BPS, $\tilde h=5$ & 0  & {\bf 23} & 0  & 3381 & {\bf 46} &202356 \\
\hline
$\frac14$-BPS, $\tilde h=6$ & {\bf 1}  & 0 &  {\bf 322}   & 0 & 29555 & 1104 \\
\hline
$\frac14$-BPS, $\tilde h=7$ & 0 & {\bf 23} & 0 & 3404 & {\bf 46} & 225011 \\
\hline
$\frac14$-BPS, $\tilde h=8$& {\bf 1} & 0 & {\bf 322} & 0 & 29855 & {\bf 1106} \\
\hline
$\frac14$-BPS, $\tilde h=9$ & 0 & {\bf 23} & 0 & {\bf 3405} & {\bf 46} & 227908 \\
\hline
$\frac14$-BPS, $\tilde h=10$ & {\bf 1} & 0 &  {\bf 322}  & 0 & 29878 & {\bf 1106} \\
\hline
$\frac14$-BPS, $\tilde h=11$ & 0 & {\bf 23} & 0 & {\bf 3405} & {\bf 46} & 228208 \\
\hline
$\frac14$-BPS, $\tilde h=12$ & {\bf 1} & 0 &  {\bf 322}  & 0 & {\bf 29879} & {\bf 1106} \\
\hline
$\frac14$-BPS, $\tilde h=13$ & 0 & {\bf 23} & 0 & {\bf 3405} & {\bf 46} & 228231 \\
\hline
$\frac14$-BPS, $\tilde h=14$ & {\bf 1} & 0 & {\bf 322} & 0 &{\bf 29879} & {\bf 1106} \\
\hline
$\frac14$-BPS, $\tilde h=15$ & 0 & {\bf 23} & 0 & {\bf 3405} & {\bf 46} & {\bf 228232} \\
\hline

\end{tabular}
\end{center}
\caption{Stabilization of degeneracies of BPS states at large $i$. Bold nonzero numbers indicate the degeneracies have already stabilized (see (\ref{eq:stabcondk3})).}
\label{tab:regulartable}
\end{table}

\begin{table}[h]
\begin{center}
\def\arraystretch{1.2}
\begin{tabular}{| c || c | c | c | c | c | c |}
\hline

&  $\check \chi_{i,\tilde h}^\ell \overline{\check\chi_{3i+5\tilde h}^s}$ & $\check\chi_{i,\tilde h}^\ell\overline{\check \chi_{3i+5\tilde h-1}^s}$ & $\check\chi_{i,\tilde h}^\ell \overline{\check\chi_{3i+5\tilde h-2}^s}$ & $\check\chi_{i,\tilde h}^\ell \overline{\check\chi_{3i+5\tilde h-3}^s}$ & $\check\chi_{i,\tilde h}^\ell\overline{\check\chi_{3i+5\tilde h-4}^s}$ & $\check\chi_{i,\tilde h}^\ell \overline{\check\chi_{3i+5\tilde h-5}^s}$\\
\hline
$\frac12$-BPS, $\tilde h=0$ & 1 & 4 & 12 & 32 &  81 & 192 \\
\hline
$\frac14$-BPS, $\tilde h=1$ & 4  & 22 & 76  & 223 & 600 & 1505\\
\hline
$\frac14$-BPS, $\tilde h=2$ & 7 & 56 & 240  & 816 & 2447 & 6702 \\
\hline
$\frac14$-BPS, $\tilde h=3$ & {\bf 8}  & 91 & 500  & 1982 & 6648 & 19973\\
\hline
$\frac14$-BPS, $\tilde h=4$ & {\bf 8}  & 112 & 769  & 3598 & 13593 & 44938 \\
\hline
$\frac14$-BPS, $\tilde h=5$ & {\bf 8} & 119 & 952  & 5218 & 22338 &81456 \\
\hline
$\frac14$-BPS, $\tilde h=6$ & {\bf 8}  &  {\bf 120} & 1036  & 6400 & 30829 & 124086 \\
\hline
$\frac14$-BPS, $\tilde h=7$ & {\bf 8} &  {\bf 120} & 1064 & 7038 & 37202 & 164185 \\
\hline
$\frac14$-BPS, $\tilde h=8$& {\bf 8} &  {\bf 120} & 1071 & 7304 & 40953 & 194680 \\
\hline
$\frac14$-BPS, $\tilde h=9$ & {\bf 8} &  {\bf 120} & {\bf 1072} & 7396 & 42752 & 213630 \\
\hline
$\frac14$-BPS, $\tilde h=10$ &{\bf 8} &  {\bf 120} & {\bf 1072} & 7424 & 43494 & 223530 \\
\hline
$\frac14$-BPS, $\tilde h=11$ & {\bf 8} &  {\bf 120} & {\bf 1072} & 7431 & 43768 & 228074 \\
\hline
$\frac14$-BPS, $\tilde h=12$ & {\bf 8} &  {\bf 120} & {\bf 1072} & {\bf 7432} & 43860 & 229984 \\
\hline
$\frac14$-BPS, $\tilde h=13$ & {\bf 8} &  {\bf 120} & {\bf 1072} & {\bf 7432} & 43888 & 230734 \\
\hline
$\frac14$-BPS, $\tilde h=14$ & {\bf 8} &  {\bf 120} & {\bf 1072} & {\bf 7432} & 43895 & 231008 \\
\hline
$\frac14$-BPS, $\tilde h=15$ & {\bf 8} &  {\bf 120} & {\bf 1072} & {\bf 7432} & {\bf 43896} & 231100 \\
\hline
$\frac14$-BPS, $\tilde h=16$ & {\bf 8} &  {\bf 120} & {\bf 1072} & {\bf 7432} & {\bf 43896} & 231128 \\
\hline
$\frac14$-BPS, $\tilde h=17$ & {\bf 8} &  {\bf 120} & {\bf 1072} & {\bf 7432} & {\bf 43896} & 231135 \\
\hline
$\frac14$-BPS, $\tilde h=18$ & {\bf 8} & {\bf 120} & {\bf 1072} & {\bf 7432} & {\bf 43896} & {\bf 231136} \\
\hline

\end{tabular}
\end{center}
\caption{Stabilization of degeneracies of BPS states at large $i$ for $T^4$. Bold numbers indicate the degeneracy has already stabilized (see (\ref{eq:stabcondt4})).}
\label{tab:regulartableT4}
\end{table}

Since the 1/4--BPS degeneracies are moduli--dependent, it would be interesting to see how the stabilization changes at other points in moduli space. Empirically it appears that at the symmetric orbifold point, there is a symmetry where we can increase the left-moving $SU(2)$ spin by $\frac12$ and the right-moving $SU(2)$ spin by $\frac32$, the same symmetry which occurs for the moduli--independent 1/2--BPS degeneracies. However, there does not appear to be a symmetry where we increase the left--moving conformal weight by $2$ and the right-moving spin by $5$. It would be interesting to explore this further, at the orbifold point or at other potentially interesting points in moduli space.  Finally, a natural question is if at large $N$ there is some enhanced symmetry beyond $\mathcal N=(4,4)$ underlying this stabilization phenomenon. See \S \ref{sec:discuss} for further discussion on this point.

\section{Discrete symmetries}\label{sec:symms}
In this section we consider the action of discrete symmetry groups on the refined BPS spectrum of CFTs in the moduli space of Sym$^N(M)$, for $M=K3$ and $M=T^4$. In \S \ref{sec:CFTsym} we present a general discussion of the action of such symmetries on the (BPS) spectrum of the CFT. In \S \ref{sec:sugrasym} we explicitly compute the action of certain discrete symmetry groups at the supergravity point of Sym$^N(K3)$ and comment on the case of Sym$^N(T^4)$.

\subsection{Conformal field theory}\label{sec:CFTsym}
We begin by considering the CFT with target space $M$. Such a NLSM has moduli space of the form \cite{Narain:1986am, Aspinwall:1994rg, Nahm:1999ps}
\be\label{eq:mod}
\mathcal M(M)= O(\Gamma^{4,n})\backslash O(4,n)/O(4)\times O(n),
\ee
where $n=20$ and $4$ for $M=K3$ and $T^4$, respectively. Supersymmetry--preserving discrete groups $G$ which arise at a given point in $\mathcal M(M)$ have been classified.
For the case of $M=K3$, these groups were classified in \cite{Gaberdiel:2011fg,Cheng:2016org} and shown to be isomorphic to four--plane--preserving subgroups of the group $Co_0$ (``Conway zero"), the group of automorphisms of the Leech lattice. For $M=T^4$, these groups were classified in \cite{Volpato:2014zla}, and similarly shown to be isomorphic to four--plane--preserving subgroups of $W^+(E_8)$, the group of even Weyl transformations of the $E_8$ root lattice.

Suppose $G$ is such a supersymmetry--preserving discrete symmetry group, occurring at a given point in the moduli space, $X\in \mathcal M(M)$. Then for each conjugacy class $g \in G$ we can define the following trace,
\be
{\cal Z}^X_g(\t,z, \nu)=\Tr_{\text{RR}}(-1)^{F_L+F_R}g q^{L_0-\frac{c}{24}}q^{\overline{L_0}-\frac{\overline c}{24}}y^{J_0} u^{\bar{J_0}},
\ee
which we interpret as a twined version of the CFT partition function. Clearly this function is both well-defined, as $G$ commutes with the $\mathcal N=(4,4)$ superconformal algebra, and dependent on $X$. Upon setting $\nu=0$ this function is simply a twined version of the EG; i.e. 
\be\label{eq:twineEG}
{\cal Z}^X_g(\t,z, 0)=Z^X_{\text{EG},g}(\t,z)=\sum_{m,\ell}c_g(m,\ell) q^m y^\ell
\ee
 and is furthermore a weak Jacobi form of weight zero and index one for a congruence subgroup of $SL(2,\mathbb Z)$ contained in $\Gamma_0(o(g))$, where $o(g)$ is the order of the group element $g$.\footnote{Note that though the EG of $T^4$ vanishes, many of these twining genera do not.} For $M=K3$, this function no longer depends on $X$ but only the conjugacy class of $g$ within the duality group $O^+(\Gamma^{4,20})$.\footnote{The ``+" arises because one needs to take into account worldsheet parity when considering symmetries which act asymmetrically on the left-- and right--moving fields of the NLSM. For the explicit symmetries we discuss in the next section, this subtlety will not play a role.} Furthermore, all such conjugacy classes \cite{Cheng:2016org} and their associated twining genera \cite{Paquette:2017gmb} have been classified. For $T^4$, a similar statement should hold with $\Gamma^{4,20}$ replaced by $\Gamma^{4,4}$, though, to the best of our knowledge, the associated twining genera have not been classified in the same sense as for $K3$. See \cite{Volpato:2014zla} for more details.

Similarly, given $X \in \mathcal M(M)$ with symmetry group $G$, for each conjugacy class $g\in G$, one can define a twined version of the HEG as
\be\label{eq:HEGg}
Z^X_{\text{HEG}, g}(\tau, z, \nu) :={\cal Z}^X_g(\t,z, \overline \t, \nu)\big|_{\bar h = \frac c{24} }=\Tr_{\text{RR}}\((-1)^{F_L+F_R}g q^{L_0-\frac{c}{24}}y^{J_0} u^{\bar{J_0}} \) \bigg|_{\bar h = \frac c{24} },
\ee
where by $\bar h = \frac c{24}$ we mean the restriction to right-moving Ramond ground states. Like the (twined) partition function, (\ref{eq:HEGg}) depends sensitively on the point in CFT moduli space and is in general difficult to compute for an arbitrary point in $\mathcal M(M)$ for $M=K3, T^4$. Moreover  upon setting $\nu=0$ it reduces to the twined EG. However, unlike the (twined) partition function or EG, in general it does not have any nice modular properties.

We would like to consider the properties of these various twined traces upon lifting to the symmetric product theory and theories in its moduli space.
Such a moduli space has the form \cite{Dijkgraaf:1998gf, Seiberg:1999xz} 
\be\label{eq:symmod}
\mathcal M(\text{Sym}^N(M)) = SO(4,n+1; \bb Z) \backslash SO(4,n+1; \bb R)/SO(4)\times SO(n+1)
\ee
where again $n=20$ and $4$ for $M=K3$ and $T^4$, respectively. Given a point $X \in \mathcal M(M)$ in the moduli space of NLSMs on $M$ with a symmetry group $G$, following \cite{Cheng:2010pq} one can lift this symmetry to an action on the BPS spectrum counted by the EG of Sym$^N(M)$ as follows,
\be\label{eq:DMVVg}
\sum_{N=1}^\infty Z_{{\rm EG},g}^{\text{Sym}^N(X)}(\t,z) p^N = \prod_{n>0,m\geq 0, \ell} \prod_{k=0}^{M-1}\frac1{(1-e^{2\pi i k/M} p^nq^my^\ell)^{\hat c_g(nm,\ell,k)}},
\ee
where $M$ is the order of the group element $g$ and the coefficients $\hat c_g$ are defined by
\be
\hat c_g(m,\ell,k) := \frac1M\sum_{j=0}^{M-1} e^{-\frac{2\pi i k j}M} c_{g^j}(m,\ell),
\ee
where $c_g(m,\ell)$ is as in equation (\ref{eq:twineEG}). Similarly, we expect that a similar formula lifts the action of the symmetry on the refined spectrum of BPS-states counted by the HEG to those given by the HEG of Sym$^N(X)$ in the following way. Let $c_g(m,\ell,\ell')$ denote the coefficients in the expansion of the twined HEG (\ref{eq:HEGg}) of $X$ given by
\be
Z^X_{\text{HEG}, g}(\tau, z, \nu)=\sum_{m,\ell,\ell'}c_g(m,\ell,\ell') q^my^\ell u^{\ell'}.
\ee
Then we can write a generating function for the action of this symmetry on the refined BPS spectrum of Sym$^N(X)$ as
\be\label{eq:DMVVhegg}
\sum_{N=1}^\infty Z_{{\rm HEG},g}^{\text{Sym}^N(X)}(\t,z,\nu) p^N = \prod_{n>0,m\geq 0, \ell,\ell'} \prod_{k=0}^{M-1}\frac1{(1-e^{2\pi i k/M} p^nq^my^\ell u^{\ell'})^{\hat c_g(n,m,\ell,\ell', k)}},
\ee
where $M$ is the order of $g$ and
\be
\hat c_g(n,m,\ell,\ell',k) \equiv \frac1M\sum_{j=0}^{M-1} e^{-\frac{2\pi i k j}M} c_{g^j}(nm,\ell,\ell').
\ee

At this point we would like to make some comments. 
\begin{enumerate}
\item The classification theorems of symmetries of $K3$ NLSMs in \cite{Gaberdiel:2011fg,Cheng:2016org} have as crucial input the form of the moduli space (\ref{eq:mod}). However, the  moduli space of Sym$^N(K3)$ (equation (\ref{eq:symmod})) is larger and thus it is conceivable that larger symmetry groups may arise at particular points in the moduli space of these NLSMs. The supersymmetry--preserving discrete symmetry groups of these NLSMs, however, have not been classified. Furthermore, in the case of the HEG, there are almost certainly symmetries at points in the moduli space of Sym$^N(K3)$ whose twined HEG does not arise as a lift of a symmetry of a $K3$ NLSM in the sense of (\ref{eq:DMVVhegg}). Similar comments apply in the case of $T^4$.
\item While equation (\ref{eq:DMVVhegg}) for the twined HEG of Sym$^N(X)$ relies on the point $X \in \mathcal M(M)$, and is thus only applicable at the orbifold point in the moduli space of  Sym$^N(X)$, equation (\ref{eq:DMVVg}) for the twined EG is much more general. More precisely, just as the EG of $X$ is independent of the point in $\mathcal M(M)$, the twined EG for a conjugacy class $g\in O^+(\Gamma^{4,20})$ is the same function for all points $X \in \mathcal M(M)$ with symmetry $g$. Furthermore, upon lifting this to a symmetry of Sym$^N(X)$, one expects (\ref{eq:DMVVg}) to apply to all points in $\mathcal M({\rm Sym}^N(X))$ with this symmetry.
\end{enumerate}
Given these comments it is clear that with our current technology, from the CFT perspective we can only compute the (twined) HEG at certain solvable points in the moduli space of Sym$^N(M)$. However, in the next section we will examine the twined HEG at the supergravity point of Sym$^N(M)$ from the bulk perspective for groups which arise as symmetries at points in the moduli space of $M$.

\subsection{Bulk supergravity}\label{sec:sugrasym}

We would like to understand the action of discrete symmetry groups on the refined BPS spectrum at the supergravity point in the moduli space of Sym$^N(M)$ when $N\to \infty$. We focus only on the spectrum of single- and multi-particle states accessible via the analysis of AdS$_3\times S^3$ supergravity as discussed in \S \ref{sec:spectrum}. Furthermore, we restrict ourselves to groups which preserve spacetime supersymmetry. At  the supergravity point in moduli space, the theory has continuous symmetry group $SO(4,n+1)$, where $n=20,4$ for $M=K3, T^4$, respectively, which is then broken to a discrete subgroup by charge quantization.\footnote{Note that the assignment of the ``degree" to the single--particle supergravity states described in \S \ref{sec:spectrum} also breaks this continuous symmetry group.} If we deform away from the supergravity locus to a generic point, it is plausible that the low-lying spectrum remains unchanged; thus it is possible that these discrete symmetries arise at some other point in the moduli space of Sym$^N(M)$. 

We will focus on the action of supersymmetry--preserving discrete subgroups of $SO(4, n+1)$ on the spectrum of refined BPS particle states. Our method will be to consider the action of discrete symmetry groups on the single-particle supergravity spectrum, and then use a formula analogous to (\ref{eq:DMVVhegg}) to lift this to an action on all multi-particle supergravity states. Consider a discrete symmetry group $G$ of $\mathcal N=(2,0)$ or $\mathcal N=(2,2)$ supergravity, and let $g\in G$ denote a conjugacy class. Given an explicit action of $g$ on the supergravity Hilbert space $\mathcal H^{(N)}_\chi$ described by equation (\ref{eq:halfmaxHilb}) or (\ref{eq:maxHilb}), where $\chi=(2,0)$ or $\chi=(2,2)$, respectively, let $c^\chi_{\text{sugra},g}$ denote the coefficients of 
\be\label{eq:twinetrace}
\sum_N \left(\Tr_{\mathcal H^{(N),{\rm single}}_\chi}g q^{L_0}y^{J_0}u^{\overline{J_0}}\right ) p^N = \sum_{N,m,\ell,\ell'} c^\chi_{\text{sugra},g}(N,m,\ell,\ell') p^Nq^m y^\ell u^{\ell'}.
\ee
Then we can lift this symmetry to an action on the multi-particle Hilbert space in the following way,
\be\label{eq:HEGtwined}
\sum_{N\geq0}\widetilde Z^\chi_{\text{HEG},N, g}(\tau, z, \nu) p^N = \prod_{n>0, m, \ell, \ell'} \prod_{k=0}^{M-1}\frac1{(1-e^{2\pi i k/M} p^n q^m y^\ell u^{\ell'})^{\hat c^\chi_g(n,m,\ell,\ell',k)}}
\ee
where, again, the tilde denotes that we are doing a supergravity computation,  $M$ is the order of $g$, and
\be
\hat c^\chi_g(n,m,\ell,\ell',k) = \frac1M\sum_{j=0}^{M-1} e^{-\frac{2\pi i k j}M} c^\chi_{\text{sugra},g^j}(n,m,\ell,\ell').
\ee
For the same reasons as discussed in \S\ref{sec:spectrum}, this computation will only match the twined HEG of the dual CFT for states with conformal weight below the threshold corresponding to black holes in the bulk. We now explicitly describe this action for discrete groups which can arise as symmetries of $\mathcal N=(2,0)$ supergravity, and briefly comment on the $\mathcal N=(2,2)$ case.
\subsubsection{Half--maximal supergravity}\label{sec:sugratwine}
In this section we consider discrete symmetries of $\mathcal N=(2,0)$ supergravity and their action on the refined spectrum. As mentioned earlier, charge quantization and assignment of a ``degree" to states in the spectrum break the continuous $SO(4,21)$ symmetry group. However, it is still possible to define an action on the spectrum for all $G\in Co_0$ which preserve a four--plane. We now describe this action explicitly.

It will be convenient to introduce the notion of a Frame shape as follows. Given a conjugacy class $g$ in $Co_0$, the Frame shape, $\pi_g$ is defined as
\be
\pi_g:= \prod_{\ell |M} \ell^{k_\ell}
\ee
where $M=o(g)$ is the order of $g$, $\ell$ runs over the positive divisors of $M$, and $k_\ell \in \mathbb Z$ are integers defined by the 24-dimensional irreducible representation of $g$ as
\be\label{eq:charpoly}
\det(t{\bf 1}_{24} - \rho_{24}(g)) = \prod_{\ell |M} (t^\ell-1)^{k_\ell}.
\ee
The Frame shape conveniently encodes the eigenvalues of $g$ in its 24-dimensional representation, as these are the 12 complex-conjugate pairs $\{\ell_{g,k},\overline{\ell_{g,}}\}$ which are the 24 roots of (\ref{eq:charpoly}).
Finally, we find it useful to define $\chi_g$ as the trace of $g$ in its 24-dimensional representation,
\be
\chi_g = \Tr_{\bf 24} g.
\ee

Equipped with these definitions, we can now describe explicitly the action of all $g\in Co_0$ on the multi-particle spectrum of $\mathcal N=(2,0)$ supergravity. At fixed degree $d$, the single-particle spectrum described in \S \ref{sec:halfmaxspec} has 20 states with charges $(d,d)$ and four other states with unique sets of charges (except for the case of $d=1$, which has three other states with unique sets of charges.) Thus, for a given four-plane-preserving $g$ and fixed degree $d\neq 1$, we take it to act on the 24 states of degree $d$ with four plus one eigenvalues on the states with unique charges and the remaining 20 eigenvalues on the 20 states of charge $(d,d)$. Similarly, in the case of $d=1$, we decompose the 24-dimensional representation of $g$ as ${\bf 24} = {\bf 1} + {\bf 23}$, and consider the 23-dimensional representation of $g$. In this case we take $g$ to act trivially on the three states with unique charges and to act with the remaining 20 eigenvalues on the 20 states of charge $(d,d)$.

With this action, the  coefficients $c^{(2,0)}_{\text{sugra}, g}$ defined in (\ref{eq:twinetrace}) are explicitly given by,
\begin{align}
\sum_{n,m,\ell,\ell'} &c^{(2,0)}_{\text{sugra}, g}(n,m,\ell,\ell') p^n q^m y^\ell u^{\ell'}  =\frac{1}{(1-q)(y-y^{-1})} \bigg(\frac{(u+u^{-1})p^2}{1-q^{1/2}yp}(y^2 q^{1/2} - 2 yq + q^{3/2}) \nn\\
&-\frac{(u+u^{-1})p^{2}}{1-q^{1/2}y^{-1}p}(y^{-2}q^{1/2}-2y^{-1}q+q^{3/2})+\frac{(u+u^{-1})p}{1-q^{1/2}yp}(y^3q-2y^2q^{3/2}+yq^2)\nonumber\\
&-\frac{(u+u^{-1})p}{1-q^{1/2}y^{-1}p}(y^{-3}q-2y^{-2}q^{3/2}+y^{-1}q^2)+\frac{(\chi_g-4)p}{1-q^{1/2}yp}(y^2q^{1/2}-2yq+q^{3/2})\nn\\
&-\frac{(\chi_g-4)p}{1-q^{1/2}y^{-1}p}(y^{-2}q^{1/2} -2y^{-1}q+q^{3/2})\bigg)+(u+u^{-1})p.
\label{eq:sugratwine}
\end{align}
Note that this equation is the same as equation (\ref{eq:sugra}), except  the occurrences of the ``20" are replaced by $\chi_g-4$. Using equation (\ref{eq:HEGtwined}) we can then compute the twined version of the supergravity HEG for all four-plane-preserving $g$ in $Co_0$. 

When we decompose into $\mathcal N=(4,4)$ superconformal characters, this twined function has similar stabilization properties to those described for 1/2-- and 1/4--BPS states in \S \ref{sec:halfstab} and \S \ref{sec:quartstab}. Define 
\be
\eta_g(\t):= q \prod_{k=1}^{24} \prod_{n=1}^\infty(1-\ell_{g,k}q^n),
\ee
where the $\ell_{g,k}$ are the 24 eigenvalues defined by the Frame shape of $\pi_g$ and we define $\ell_{g,k+12}:=\overline{\ell_{g,k}}$. Then the twined version of equation (\ref{eq:K3stab1}) is given by
\be\label{eq:twine1}
\sum_{k=0}^{\infty} c_g^{i, 3i-2k} q^k =q(1-q)^2(1-q^2)(1-q^3)\frac{1}{\eta_g(\t)}.
\ee
Similarly, we introduce the following definitions.\footnote{Note that in our notation, $H_{1}(\t) = H(\t)^{24}$, rather than $H(\t)$ (and similarly with other functions).}
\begin{align}
H_g(\t)&:=  \prod_{k=1}^{24} \prod_{n=1}^\infty\frac{1}{(1-\ell_{g,k}q^{5n-3})(1-\ell_{g,k}q^{5n-2})}\nn\\
G_g(\t)&:=  \prod_{k=1}^{24} \prod_{n=1}^\infty\frac{1}{(1-\ell_{g,k}q^{5n-1})(1-\ell_{g,k}q^{5n-4})}\nn\\
R_g(\t)&:=  q^{\frac{24}5}\frac{H_g(\t)}{G_g(\t)}\nn\\
\beta_g(\t)&:= \prod_{k=1}^{24} \prod_{n=1}^{\infty} \frac{(1+\ell_{g,k}q^n)^{\frac{n^2}{10}}}{(1-\ell_{g,k}q^n)^{\frac{n^2}{10}}}.
\end{align}
Furthermore, let $\ell_{-g,k}:=-\ell_{g,k}$, such that, e.g.,
\be
\eta_{-g}(\t):=q \prod_{k=1}^{24} \prod_{n=1}^\infty(1+\ell_{g,k}q^n).
\ee
Then the twined version of equation (\ref{eq:ansdiffk3}) is given by
\be
\sum_{k'} \Big | c_g^{i, \tilde h, 3i+5\tilde h-k'} - c_g^{i, \tilde h+1, 3i+5(\tilde h+1)-k'}\Big | q^{k'} = \frac{q^{\frac65}(1-q)^2(1-q^2)}{R(\tau)} \frac{H_g(\tau)}{\eta_g(\tau)}
\label{eq:asdf}
\ee
and the twined version of equation (\ref{eq:ansansdiffsk3}) is given by
\be
\sum_{k'}  \(c_g^{i, \tilde h, 3i+5\tilde h-k'} + c_g^{i, \tilde h+1, 3i+5(\tilde h+1)-k'}\) q^{k'} = \psi(\t)H_g(\t)\beta_g(\t)\frac{R_{-g}(\t)^{\frac25}}{R_g(\t)^{\frac25}}\left (\frac{\eta_{-g}(2\t)}{\eta_g(2\t)\eta_g(\t)\eta_{-g}(\t)}\right )^{\frac12},
\label{eq:asdfgh}
\ee
where $\psi(\t)$ is as defined in equation (\ref{eq:psidef}).

As we have discussed, these twinings are defined for all conjugacy classes $g \in Co_0$ which preserve a four--plane. It follows that equations (\ref{eq:twine1}), (\ref{eq:asdf}), and (\ref{eq:asdfgh}) admit decompositions into characters of four--plane--preserving subgroups $G \in Co_0$. However, one may wonder, given, e.g., the observation in \cite{Benjamin:2016pil}, whether in the large $N$ limit (or even for $N >1$) there is any role for larger subgroups of $Co_0$, i.e. subgroups $G \in Co_0$ which fix a subspace of dimension less than four. There are a couple of reasons one might ask this question. Firstly, the moduli space of NLSMs on Sym$^N(K3)$ is larger than than that of $K3$ NLSMs, and the symmetry groups haven't been classified. Therefore, there may be a role for larger discrete symmetry groups in the case of $N>1$. Secondly, there are a number of subgroups of $Co_0$ which preserve, e.g., a two-- or three--plane in the 24-dimensional representation where nevertheless, each individual conjugacy class preserves a four--plane and thus the action of each conjugacy class on the single--particle supergravity spectrum is well--defined. For example, the Mathieu groups $M_{22}$ and $M_{11}$ each preserve a three--plane, however, each individual element of these groups preserves a four--plane.

With this motivation in mind, in Appendix \ref{sec:decomp}, we use (\ref{eq:asdf}) and (\ref{eq:asdfgh}) to decompose the coefficients in (\ref{eq:ansdiffk3}) and (\ref{eq:ansansdiffsk3}) respectively into representations of the sporadic groups $M_{11}$ and $M_{22}$. In fact, in the large $N$ limit, any coefficient in the $\mathcal N=(4,4)$ decomposition of the supergravity BPS spectrum given in equation (\ref{eq:decompk3}) should admit a (virtual) decomposition into representations of these groups. Perhaps surprisingly, we find that at large $N$ every coefficient in (\ref{eq:decompk3}) can be decomposed into honest (non--virtual) $M_{22}$ representations. In Table \ref{tab:toro}, we give an example of this by providing a decomposition of the terms in (\ref{eq:sushi}) into $M_{22}$ representations. We expect a similar statement to hold for $M_{11}$, and most likely other three--plane--preserving subgroups of $Co_0$.

Finally, we briefly comment on the twinings of equation (\ref{eq:hodgeK3}), the generating function of refined 1/2--BPS degeneracies (which holds for any point in the moduli space of Sym$^N(K3)$). Such twinings were first conjectured in \cite{Cheng:2015kha} for all four--plane--preserving conjugacy classes of $Co_0$. Note that we can reproduce these conjectural twinings explicitly from the action of such symmetries on the supergravity Hilbert space described in this section. Furthermore, we find that for all $N>1$, the decomposition of the refined 1/2--BPS degeneracies into $M_{22}$ representations is non-virtual, suggesting this group may play a role in the symmetries of the refined 1/2--BPS spectrum of symmetric product theories.

\subsubsection{Maximal supergravity}
Given the explicit description of four--plane--preserving conjugacy classes of $W^+(E_8)$ in \cite{Volpato:2014zla}, it should be possible to do an analysis for $\mathcal N=(2,2)$ supergravity similar to that of the previous section. We expect that we would be able to define a twined version of the stabilized degeneracies of equation (\ref{eq:t4sumwoo}) at large $N$ for all $g \in W^+(E_8)$ which preserve a four--plane. An interesting question is whether one can find evidence for a role of larger subgroups of $W^+(E_8)$ when $N>1$, as we found in the case of half--maximal supergravity in the previous section. We leave a detailed analysis of the action of these groups on the refined $\mathcal N=(2,2)$ spectrum to future work.

\section{Discussion}
\label{sec:discuss}

In this paper we explored the BPS spectrum of the symmetric orbifold of $K3$ and $T^4$ deformed to the supergravity point in moduli space, including refinement under both left-- and right--moving $SU(2)_R$ charges. We found interesting symmetry properties both of the 1/2--BPS and 1/4--BPS spectrum that potentially suggest deeper structure.

In \S\ref{sec:theanswer} we find that the degeneracies of BPS states at the supergravity point satisfy an interesting stabilization phenomenon. In particular, if the $SU(2)_R$ spin on the right is large enough, we find that there are two minimal operations which leave the degeneracy unchanged: (a) increase the spin on the left by $\frac12$ and the spin on the right by $\frac32$; or (b) increase the conformal weight on the left by $2$ and the spin on the right by $5$. (For the case of maximal supergravity, as the parity of the left--moving conformal weight is irrelevant, the minimal version of b) is to increase the conformal weight on the left by $1$ and the spin on the right by $\frac52$.)

This seems to suggest a possible hidden symmetry of the theory that acts on the left-- and right--movers in precisely this fashion. Furthermore, the generating functions for the BPS-state degeneracies at large right-moving spin (given in (\ref{eq:ansdiffk3}), (\ref{eq:ansansdiffsk3}), and (\ref{eq:t4sumwoo})) all miraculously involve Rogers-Ramanujan functions. These functions have shown up in various contexts in physics, for instance as the characters of the two primary operators of the non-unitary $(2,5)$ Virasoro minimal model \cite{Yellow}; it would be interesting to have an explanation why they appear here.

Secondly, in \S\ref{sec:symms} we discuss the action of supersymmetry--preserving discrete symmetry groups on the refined BPS spectrum of Sym$^N(M)$ for $M=K3$ and $T^4$. We explicitly derive an action for all four--plane--preserving subgroups of $Co_0$ on the multi--particle spectrum of $\mathcal N=(2,0)$ supergravity at large $N$, and derive an analytic formula for the twined supergravity HEG in the regime where the degeneracies have stabilized. Furthermore, we comment on the potential action of larger (i.e. three--plane--preserving) symmetry groups on the large--$N$ 1/4--BPS spectrum and provide (non-virtual) decompositions of the stabilized degeneracies into irreducible representations of the sporadic groups $M_{22}$ and $M_{11}$ as evidence.

Finally we end with a list of additional questions which we find interesting.

\begin{itemize}
\item What explains the near modular properties of the counting functions for stabilized BPS degeneracies; i.e. equations (\ref{eq:K3stab1}) and (\ref{eq:T4stab1}) for 1/2--BPS degeneracies and (\ref{eq:ansdiffk3}), (\ref{eq:ansansdiffsk3}), and (\ref{eq:t4sumwoo}) for 1/4--BPS degeneracies? Note that the would-be ``modular parameter" in these equations is not the usual $\tau$ which couples to the left--moving conformal weight but a parameter which couples to the invariant combination of left-- and right--moving quantum numbers. ($k=3i-j$ in the 1/2--BPS case and $k'=3i+5\tilde h -j$ in the 1/4--BPS case.)

\item Is there a hidden symmetry or deeper structure of the CFT that would explain the stabilization phenomena found in \S\ref{sec:theanswer}? Note that this cannot simply be an enhancement of the chiral algebra to something larger than $\mathcal N=4$ as the symmetry couples the left-- and right--moving quantum numbers. 

\item Does the presence of Rogers-Ramanujan functions indicate any connections to affine Lie algebras? For example, can (\ref{eq:ansdiffk3}), (\ref{eq:ansansdiffsk3}), and (\ref{eq:t4sumwoo}) be explained by the presence of a symmetry acting on an auxiliary Hilbert space, similar to \cite{Nakajima}? 

\item Do similar stabilization properties hold away from the supergravity locus? For instance, what happens at the orbifold point, or at a point in moduli space close to the orbifold point? Note that, as mentioned in \S \ref{sec:quartstab}, it appears that property (a) above holds at the orbifold point. Do we see similar structure at the point studied in \cite{Raju:2007uj}, for instance?

\item What happens if we include states that are dual to black holes\footnote{Or black rings, see e.g. \cite{Emparan:2006mm}.} as opposed to just supergravity KK modes? Can we derive the full HEG including these states? Do these degeneracies have the same stabilization properties?

\item Do similar stabilization phenomena hold for the refined BPS spectra other theories dual to AdS$_3$ supergravity, such as theories with, e.g., large $\mathcal N=4$ superconformal symmetry \cite{Gukov:2004ym,Eberhardt:2017fsi,Eberhardt:2017pty} or theories which arise from the MSW string \cite{Maldacena:1997de}?

\item Is there any relation between our results in \S \ref{sec:theanswer} and the stabilization phenomena observed in \cite{Wong:2017wfe}, from studying orbifolds of NLSMs whose target space is an ADE surface singularity? Is there a generalization of the connection discussed in \cite{Cheng:2014zpa} between the EG of ADE singularities, the $K3$ EG, and umbral moonshine \cite{Cheng:2013wca} to the refined BPS spectrum?

\item What is the classification of supersymmetry--preserving discrete symmetry groups of NLSMs in the moduli space of Sym$^N(M)$?  For $M=K3$ do subgroups of $Co_0$ which preserve less than a four--plane play a role?

\item In \cite{Cheng:2015kha} twinings of refined 1/2--BPS states counted by (\ref{eq:hodgeK3}) under four--plane--preserving elements of $Co_0$  were proposed to be connected to symmetries of the chiral ring of an auxiliary conformal field theory with symmetry group $Co_0$ \cite{Duncan:2014eha}. In \cite{Cheng:2014owa} it was described how this conformal field theory naturally furnishes modules for a number of sporadic groups, including the Mathieu groups $M_{22}$ and $M_{11}$. Is there any connection between the action of these groups on the states of this auxiliary conformal field theory and the $M_{22}$ and $M_{11}$ discussed in \S \ref{sec:sugratwine} which may play a role in the refined 1/4--BPS supergravity spectrum?

\item The results of \cite{Cheng:2015kha} regarding the twined 1/2--BPS spectrum have a natural geometric interpretation: they capture $g$-equivariant reduced refined Gopakumar--Vafa invariants of a $K3$ surface (c.f. Conjecture 3 of \cite{Cheng:2015kha}.) Furthermore, in \cite{Kachru:2016igs} it is proposed that the refined 1/4--BPS spectrum captured by the HEG also has a geometric interpretation. A natural question is then whether our results for the stabilization of these (twined) 1/4--BPS degeneracies at large $N$ can be understood in a geometric context.

\item Do we see any hint of mock modularity when we further reduce to AdS$_2 \times S^2$ (see e.g. \cite{Dabholkar:2012nd})?\footnote{We thank the anonymous referee for pointing out this interesting question.}

\end{itemize}

{\centerline{\bf{Acknowledgements}}

\medskip

It is a pleasure to thank Miranda Cheng, Rajesh Gopakumar, Shamit Kachru, Aaron Landesman, Hannah Larson, Shiraz Minwalla, Suvrat Raju, Shu-Heng Shao, Arnav Tripathy, Roberto Volpato, and Kenny Wong for relevant discussions, and Ethan Dyer for irrelevant discussions. We would also like to thank Shamit Kachru and Suvrat Raju for helpful comments on a draft. NB thanks the ICTS for their hospitality during which some of this work was completed, and is supported by a Stanford Graduate Fellowship and an NSF Graduate Fellowship. SMH is supported by a Harvard University Golub Fellowship in the Physical Sciences and DOE grant DE-SC0007870. She acknowledges the kind hospitality of the
Aspen Center for Physics, which is supported by NSF grant PHY-1066293,
as this was being completed.

\appendix

\section{Special Functions and Characters}
\label{app:thetachar}

In this appendix, we give definitions to various special functions used throughout the paper, including character formulas. In this section, and throughout the text, we define
\be
q = e^{2\pi i \tau}, ~~~~y = e^{2\pi i z}.
\ee

We define the following Jacobi theta functions
\begin{align}
\theta_1(\tau, z) &= -i q^{\frac18} (y^\frac12-y^{-\frac12})\prod_{n=1}^{\infty} (1-q^n)(1-yq^n)(1-y^{-1}q^{n}) \nn \\
\theta_2(\tau, z) &= q^\frac18 (y^\frac12 + y^{-\frac12})\prod_{n=1}^{\infty} (1-q^n)(1+yq^n)(1+y^{-1}q^{n}) \nn \\
\theta_3(\tau, z) &= \prod_{n=1}^{\infty} (1-q^n)(1+yq^{n-\frac12})(1+y^{-1}q^{n-\frac12}) \nn \\
\theta_4(\tau, z) &= \prod_{n=1}^{\infty} (1-q^n)(1-yq^{n-\frac12})(1-y^{-1}q^{n-\frac12})
\end{align}
as well as the Dedekind eta function
\be
\eta(\tau) = q^{\frac{1}{24}}\prod_{n=1}^{\infty} (1-q^n)
\ee
and the modular discriminant 
\be
\Delta(\tau) = \eta(\tau)^{24}.
\ee
Finally we define the Rogers-Ramanujan functions $G(\tau), H(\tau)$ as
\begin{align}
G(\tau) &= \prod_{n=1}^{\infty} \frac1{(1-q^{5n-4})(1-q^{5n-1})}\nn\\
H(\tau) &= \prod_{n=1}^{\infty} \frac1{(1-q^{5n-3})(1-q^{5n-2})}.
\end{align}
and their ratio as
\be
R(\t)= q^{\frac15}\frac{H(\tau)}{G(\tau)}.
\ee
Note that the function $R(\tau)$ is the Hauptmodul of the congruence subgroup $\Gamma(5)$ (see e.g. \cite{RR1, RR2}), defined as
\be
\Gamma(5)=\left\{\begin{pmatrix} a&b\\c&d \end{pmatrix} \el SL(2,\bb Z) ~~~~ a, d \equiv \pm 1~(\text{mod}~5), ~~b, c \equiv 0~(\text{mod}~5)\right\}.
\ee

\subsection{Small $\mc{N}=4$ Characters}\label{app:smallchars}

In this section we present the characters for the small $\mc{N}=4$ superconformal algebra, following \cite{Eguchi:1987sm, Eguchi:1987wf}. The representations of the algebra are labelled by the conformal weight and the spin of their highest weight state. The representations  come in two classes: short (or BPS) and long (or non-BPS); the short multiplets satisfy that, in the NS sector, the conformal weight is equal to the spin; the long multiplets have the conformal weight greater than the spin. We define the character of a representation as
\be
\chi(\tau, z) = \Tr_{\mathcal H}\((-1)^{J_0} q^{L_0} y^{J_0}\), 
\ee
where $\mathcal H$ is either the Ramond or NS Hilbert space and by $J_0$ we mean the Cartan of the $SU(2)$ current algebra which is part of the $\mathcal N=4$ superconformal algebra.
By convention we will label each representation in the following way. For each highest weight representation, let $h, j$ denote the eigenvalues of $L_0,J_0$ in the NS sector, respectively, where $J_0$ is the Cartan of the $SU(2)$ of the $\mathcal N=4$. Furthermore, let $\tilde h \equiv h-j/2$ be the difference between the conformal weight and the spin of the highest weight state. If the central charge is given by $c=6m$, there are $m+1$  short representations with quantum numbers $\tilde h =0; j=0, 1, \ldots, m$, and $m$ families of long representations with quantum numbers given by  $j= 0, 1, \ldots m-1$, and $\tilde h>0$. 

The NS sector characters are then given by
\begin{align}
\chi^{s, \text{NS}}_{j;m}(\tau,z) &= q^{j/2} (-1)^{j} \left(\frac{-i q^{1/4} \theta_4(\tau, z)^2}{\theta_1(\tau, 2z)\eta(\tau)^3}\right) \nonumber\\ &~~~\sum_{k\in\mathbb{Z}} q^{(m+1)k^2+(j+1)k}\left(\frac{y^{2(m+1)k+j+1}}{(1-yq^{k+\frac12})^2} - \frac{y^{-2(m+1)k-j-1}}{(1-y^{-1}q^{k+\frac12})^2}\right) \nonumber \\
\chi^{\ell, \text{NS}}_{j, \tilde h;m}(\tau,z) &= q^{j/2+\tilde h}(-1)^{j} \left(\frac{-i q^{1/4} \theta_4(\tau,z)^2}{\theta_1(\tau,2z)\eta(\tau)^3}\right) \nonumber\\& ~~~\sum_{k\in\mathbb{Z}}q^{(m+1)k^2+(j+1)k}\left(y^{2(m+1)k+j+1}-y^{-2(m+1)k-j-1}\right).
\label{eq:hawks}
\end{align}

We can obtain the Ramond-sector characters by spectral flowing by $1/2$ unit\footnote{For convenience, in (\ref{eq:hawks}) and (\ref{eq:horses}), we define the characters with a relative shift so that both the NS vacuum and R vacua characters begin at $q^0$. In other words, the NS characters are defined as $\Tr_{
\text{NS}}\((-1)^F q^{L_0} y^{J_0}\)$ and the R characters are defined as $\Tr_\text{R}\((-1)^F q^{L_0-\frac{m}4} y^{J_0}\)$. Note that the NS character differs from the usual definition of characters by a factor of $q^{-c/24}$. }:
\begin{align}
\chi^{s, \text{R}}_{j;m}(\tau,z) &= (-1)^{j+m} \frac{i \theta_1(\tau, z)^2}{\theta_1(\tau, 2z)\eta(\tau)^3} \sum_{k\in\mathbb{Z}} \frac{q^{(m+1)k^2+k}y^{2(m+1)k+1}}{(1-yq^k)^2}\left(y^{m-j+1} q^{k(m-j+1)} - y^{-(m-j+1)}q^{-k(m-j+1)}\right) \nonumber \\
\chi^{\ell, \text{R}}_{j, \tilde h;m}(\tau,z) &= q^{\tilde h}(-1)^{j+m} \frac{i \theta_1(\tau,z)^2}{\theta_1(\tau,2z)\eta(\tau)^3} \sum_{k\in\mathbb{Z}}q^{(m+1)k^2}y^{2(m+1)k}\left(q^{k(m-j)}y^{m-j}-q^{-k(m-j)}y^{-(m-j)}\right).
\label{eq:horses}
\end{align}
In the above, we continue to label the representations by the eigenvalues $\tilde h, j$ of the highest weight state in the NS sector.

\subsection{Contracted Large $\mc{N}=4$ Characters}
\label{app:largen4}

In this paper we also consider the contracted large $\mc{N}=4$ superconformal algebra. In a 2d SCFT with target $T^4$, the superconformal symmetry  on the worldsheet is a Wigner contraction of the large $\mc{N}=4$ superconformal algebra \cite{Sevrin:1988ew, Maldacena:1999bp}. The large $\mc{N}=4$ algebra, $A_{k^{+}, k^{-}}$ is labelled by two parameters $k^{+}, k^{-}$, with central charge given by 
\be
c= \frac{6k^{+} k^{-}}{k^{+}+k^{-}}.
\ee
The characters for this algebra were computed in \cite{Petersen:1989zz, Petersen:1989pp}; to obtain the Wigner contraction we take $k^{-} \rightarrow \infty$, and get (relabelling $k^{+}$ as $m$)
\be
c=6m.
\ee
As in the small $\mc{N}=4$ algebra, there are two types of characters: short and long, which we will label by the $L_0$ and $J_0$ eigenvalues of their highest weight state in the NS sector ($h$ and $j$ respectively). Here $J_0$ is the Cartan of the $SU(2)$ R-symmetry. The short representations satisfy $h=\frac{j}2$, and so are labelled by one number $j$ which ranges from $j=0, 1, \ldots m-1$. The long characters are labelled by two numbers $j, \tilde h$ such that $h=\frac{j}2 + \tilde k$; this ranges from $j=0, 1, \ldots m-2$, $\tilde h >0$. The characters are then given by
\begin{align}
\check \chi^{s,S}_{j; m}(\tau,z) &= \frac{q^{1/8} L^S(\tau,z)}{\eta(\tau)^3} \chi^{s,S}_{j; m-1}(\tau,z)  \nn\\
\check \chi^{\ell,S}_{j, \tilde h; m}(\tau,z) &= \frac{q^{1/8} L^S(\tau,z)}{\eta(\tau)^3} \chi^{\ell,S}_{j, \tilde h; m-1}(\tau,z) 
\label{eq:largechar}
\end{align}
where $S$ denotes the sector (either $S=\text{NS}$ or $S=\text{R}$), $\chi^{s,S}_{j; m-1}$ and $\chi^{\ell,S}_{j, \tilde h; m-1}$ are the small $\mc{N}=4$ characters at central charge $c=6(m-1)$, and $L^S(\tau,z)$ is the unique long $\mc{N}=4$ character at $c=6$ starting at $q^0$:
\be
L^S(\tau,z) = \chi^{\ell,S}_{0,0; 1}(\tau,z).
\ee
Specifically, we have
\begin{align}
L^{\text{NS}}(\tau, z) &= \chi^{\ell, \text{NS}}_{0,0;1}(\tau, z) \nn\\
&= \left(\frac{-i q^{1/4} \theta_4(\tau,z)^2}{\theta_1(\tau,2z)\eta(\tau)^3}\right) \sum_{k\in\mathbb{Z}}q^{k(2k+1)}\left(y^{4k+1}-y^{-4k-1}\right) \nn\\
&= 1 - (2y^{-1} + 2y)q^{\frac12} + (y^{-2} + 5 + y^2)q - (6y^{-1}+6y)q^{\frac32} + \mc{O}(q^2)
\end{align}
and
\begin{align}
L^{\text{R}}(\tau, z) &= \chi^{\ell, \text{R}}_{0,0;1}(\tau, z) \nn\\
&= \frac{-i \theta_1(\tau,z)^2}{\theta_1(\tau,2z)\eta(\tau)^3} \sum_{k\in\mathbb{Z}}\(q^{k(2k+1)}y^{4k+1} - q^{k(2k-1)}y^{4k-1}\) \nn\\
&= (-y^{-1}+2-y) + (2y^{-2} -5y^{-1}+6-5y+2y^2)q + \mc{O}(q^2).
\end{align}

\section{Derivation of Stabilization}
\label{app:derivation}

In this appendix, we derive the stabilization phenomena discussed in \S\ref{sec:theanswer}. In the following derivations, we make repeated use of the following fact. Suppose we are given an equation of the form
\be\label{eq:rel1}
\sum_{n=0}^\infty p^n f_n(\vec x) = \(\frac{1}{1-p}\) g(p, \vec x),
\ee
where $\vec x$ is some set of additional variables, and we would like to extract the behavior of the function $f_n(\vec x)$ as $n\to \infty$. Furthermore, suppose that $g(p,\vec x)$ has the following expansion in terms of $p$,
\be\nonumber
g(p,\vec x) \sim 1 + \sum_{m=1}^\infty \alpha_m(\vec x) p^m.
\ee
Noting that the RHS of (\ref{eq:rel1}) has the form
\be\nonumber
(1 + p + p^2 + \ldots) g(p,\vec x)= 1 + \sum_{n=1}^\infty p^n \left (1 + \sum_{m=1}^n \alpha_m(\vec x)\right),
\ee
we find that 
\be
\lim_{n\to \infty}f_n(\vec x)= g(1,\vec x).
\label{eq:rel2}
\ee
We will make repeated use of (\ref{eq:rel2}) below.

Finally we note that although we believe the stabilization phenomena hold whenever (\ref{eq:stabcondk3}) or (\ref{eq:stabcondt4}) is satisfied (for the case of half-maximal and maximal supergravity respectively), in this appendix we will only derive the stabilization when we take the limit $i\rightarrow\infty$ and then $\tilde h \rightarrow\infty$. It would be interesting to derive the stricter conditions (including when we change the order of limits of $i$ and $h$).

\subsection{Half--maximal supergravity}

We begin with an analysis of the refined 1/2-- and 1/4--BPS spectrum of the supergravity limit of Type IIB compactification on AdS$_3\times S^3 \times K3$.
\subsubsection{Half--BPS states}\label{sec:derivK3half}

First let's derive the large $N$ behavior of the refined degeneracies of half-BPS states. These are moduli-independent. The generating function for the half-BPS states given in equation (\ref{eq:hodgeK3}) can be written
\be\label{eq:zip}
\sum_{N=0}^\infty \sum_{i,j=0}^N c_N^{i,j} p^N y_N^{[i]} u_N^{[j]} = \prod_{n=1}^\infty \frac{1}{(1-p^n)^{20}(1-p^n uy)(1-p^n u^{-1} y)(1-p^n u y^{-1}) (1-p^n u^{-1}y^{-1})},
\ee
where $y_N^{[i]}$ is the $q^0$ term of the short $\mc N=4$ Ramond character $\chi^{s,R}_{i;N}$ at central charge $6N$, i.e.
\be\label{eq:Rchars}
y_N^{[i]} = (-1)^{N+i} \(y^{-(N-i)} + y^{-(N-i)+2} + \ldots + y^{N-i}\)= (-1)^{N+i}\(\frac{y^{N-i+1}-y^{-(N-i+1)}}{y-y^{-1}}\).
\ee
We are interested in the behavior of the coefficients $c_N^{i,j}$, which count the number of half-BPS states of CFTs in the moduli space of Sym$^N(K3)$ with $J_0, \overline {J_0}$ eigenvalues $i$ and $j$, respectively.
Let's calculate the large central-charge limit of (\ref{eq:zip}); i.e. the limit
\be
\lim_{N\to \infty}\sum_{i,j=0}^N c_N^{i,j} y_N^{[i]} u_N^{[j]}.
\ee
 First we plug (\ref{eq:Rchars}) into (\ref{eq:zip}), divide both sides by $yu$, and define a new variable $\tilde p = pyu$ to rewrite (\ref{eq:zip}) as
\begin{align}
\sum_{N=0}^{\infty} \sum_{i,j=0}^N &(-1)^{i+j}c_N^{i,j} \tilde p^N \( y^{-i} u^{-j} - y^{-i} u^{-2N+j-2} - y^{-2N+i-2}u^{-j} + y^{-2N+i-2}u^{-2N+j-2}\) \nn\\ 
&= \frac{(1-y^{-2})(1-u^{-2})}{1-\tilde p} \Bigg(\prod_{n=1}^\infty \frac{1}{(1-\tilde p^n u^{-n} y^{-n})^{20}(1-\tilde p^{n+1} u^{-n} y^{-n} )} \nn \\ 
& \phantom{aaa} \frac{1}{(1-\tilde p^n u^{-n-1} y^{-n+1}) (1-\tilde p^n u^{-n+1} y^{-n-1})(1-\tilde p^n u^{-n-1} y^{-n-1}) }\Bigg).
\end{align}
We note that this is an expansion in $\tilde p$ which has the form (\ref{eq:rel1}), so to take the large $N$ limit, we use (\ref{eq:rel2}) to get
\begin{align}
&\lim_{N\rightarrow\infty} \sum_{i,j=0}^{N} (-1)^{i+j}c^{i,j}_N \( y^{-i} u^{-j} - y^{-i} u^{-2N+j-2} - y^{-2N+i-2}u^{-j} + y^{-2N+i-2}u^{-2N+j-2}\) \nn\\  
&= (1-y^{-2})(1-u^{-2}) \prod_{n=1}^\infty \frac{1}{(1-u^{-n} y^{-n})^{21}(1- u^{-n-1} y^{-n+1}) (1-u^{-n+1} y^{-n-1})(1- u^{-n-1} y^{-n-1}) }.
\end{align}
Finally, we are interested in the coefficient $c_N^{i, 3i-2k}$ in the large $i$ limit (after we have taken $N$ large). Redefining $\tilde y = y^{-1}u^{-3}$ and again using (\ref{eq:rel1}) and (\ref{eq:rel2}) for the variable $\tilde y$, we find that
\begin{align}
\lim_{i\rightarrow \infty} \lim_{N\rightarrow\infty} \sum_{i,j=0}^N (-1)^{i+j} c_N^{i,j} u^{3i-j} &= \frac{(1-u^6)(1-u^{-2})}{1-u^{-2}}\prod_{n=1}^{\infty} \frac{1}{(1-u^{2n})^{21}(1-u^{2n})(1-u^{2n+4})(1-u^{2n+2})} \nn\\
&= (1-u^2)^2(1-u^4)(1-u^6)\prod_{n=1}^{\infty} \frac{1}{(1-u^{2n})^{24}}.
\label{eq:aaaaaa}
\end{align}
In the RHS of (\ref{eq:aaaaaa}), only even powers of $u$ show up, so on the LHS, $i+j$ must be even, so we can drop the $(-1)^{i+j}$ term. Thus we show that at large $N$ and $i$ (where we take $N$ to infinity first), $c_N^{i,3i-2k}$ is independent of $i, N$, and is given by the generating function of equation (\ref{eq:K3stab1}).

\subsubsection{Quarter--BPS states}\label{sec:derivK3quarter}
There is a similar phenomenon for 1/4--BPS states. The generating function of multiparticle 1/4--BPS states  given by equation (\ref{eq:fullheg}) can be decomposed into characters as\footnote{Note that strictly speaking the RHS of (\ref{eq:above}) also gets contributions from 1/2--BPS states (see equation (\ref{eq:decompk3})); however we will only examine the 1/4--BPS states in this subsection.}
\be
\sum_{N=0}^\infty\sum_{i,\tilde h, j} p^N c_N^{i, \tilde h, j} \chi^{\ell,\text{NS}}_{i, \tilde h;N}(\tau,z) \overline{\chi^{s,\text{R}}_{j;N}}(\nu) = \prod_{n>0, m, \ell, \ell'} \frac{1}{(1-p^n q^m y^\ell u^{\ell'})^{c^{(2,0)}_{\text{sugra}}(n,m,\ell,\ell')}},
\label{eq:above}
\ee
where $c^{(2,0)}_{\text{sugra}}$ is defined in (\ref{eq:sugra}). In this formula $c_N^{i, \tilde h, j}$ denotes the degeneracy quarter-BPS represenations of the dual CFT whose highest weight state has NS-NS eigenvalues of $(L_0, J_0, \overline{J_0})$ given by $(i/2 + \tilde h, i, j)$. The $\mc{N}=4$ characters (given in (\ref{eq:hawks})) simplify at large central charge in the following way
\begin{align}
\chi^{s,\text{NS}}_{j;\infty}(\tau,z) &= q^{j/2} (-1)^{j} \(\frac{-i q^{1/4} \theta_4(\tau, z)^2}{\theta_1(\tau, 2z)\eta(\tau)^3}\)\(\frac{y^{j+1}}{(1-yq^{\frac12})^2} - \frac{y^{-j-1}}{(1-y^{-1}q^{\frac12})^2}\), \nonumber \\
\chi^{\ell,\text{NS}}_{j, \tilde h;\infty}(\tau,z) &= q^{j/2+\tilde h}(-1)^{j} \left(\frac{-i q^{1/4} \theta_4(\tau,z)^2}{\theta_1(\tau,2z)\eta(\tau)^3}\right)\left(y^{j+1}-y^{-j-1}\right).
\label{eq:hawksatlargeC}
\end{align}

Using (\ref{eq:hawksatlargeC}), we can write (\ref{eq:above}) as,
\begin{align}
\sum_{N=0}^\infty\sum_{i,\tilde h,j}(-1)^{i+j} &p^Nq^{\frac i2+\tilde h} c_N^{i, \tilde h, j} (y^{i+1}-y^{-(i+1)})\(\frac{-i q^{1/4} \th_4(\tau,z)^2}{\th_1(\tau,2z)\eta(\tau)^3}\)\(\frac{u^{N-j+1}-u^{-(N-j+1)}}{u-u^{-1}}\) = \nn \\
&\frac{1}{(1-pu)(1-pu^{-1})}\prod_{n>1,m,\ell,\ell'}\frac{1}{(1-p^n q^m y^\ell u^{\ell'})^{c^{(2,0)}_{\text{sugra}}(n,m,\ell,\ell')}}.
\end{align}
Note that this is only valid at large $N$, so we need to take the $N \rightarrow \infty$ limit of (\ref{eq:above}). We redefine $\tilde p = pu$, remove a factor of $(1-\tilde p)^{-1}$ and use (\ref{eq:rel2}) to get,
\be
\sum_{i,\tilde h,j}(-1)^{i+j} q^{\frac i2+\tilde h} c^{i, \tilde h, j} \( y^{i+1} - y^{-(i+1)}\) u^{-j+1} = \frac{i u \th_1(\tau,2z)\eta(\tau)^3}{\th_4(\tau,z)^2 q^{1/4}}\prod_{m, \ell, \ell'} \frac{1}{(1-q^m y^\ell u^{\ell'})^{c^{(2,0)}(m, \ell, \ell')}}
\label{eq:simp}
\ee
where we have dropped the subscript of $N$ on $c^{i, \tilde h, j}$ to indicate that we have taken the limit as $N\to\infty$ and for convenience, we have defined a new set of coefficients by
\begin{align}
\sum_{m, \ell, \ell'} c^{(2,0)}(m, \ell, \ell') q^m y^\ell u^{\ell'} &= \(f^{(2,0)}(u^{-1}, q, y, u) - (1+u^{-2})\) \nn\\
&= \(\frac{\sqrt{q}}{(1-q)u^2 y(u-\sqrt q y)(uy-\sqrt q)}\) \big((uy + 21u^3y + uy^3 + 21u^3y^3) \nn\\ &~~~+ (u^2+u^4-y^2-  2uy^2-20u^2y^2-42u^3y^2+u^4y^2+u^2y^4+u^4y^4)q^{1/2} \nn\\&~~~ -(uy+2u^2y+u^3y+2u^4y+uy^3+2u^2y^3+u^3y^3+2u^4y^3)q \nn\\&~~~+ (y^2 + 2uy^2 + 22u^2y^2 + 2u^3y^2 + u^4y^2)q^{3/2}\big) \nn\\
&\equiv g^{(2,0)}(q,y,u),
\end{align}
and 
$f^{(2,0)}(p, q, y, u)$ is as in (\ref{eq:sugra}).

Now let's take the large $i$ limit of equation (\ref{eq:simp}), corresponding to taking the left-moving spin large. We will show that the degeneracies $c^{i, \tilde h, j}$ only depend on the combination $3i + 5\tilde h - j$ (as well as the parity of $\tilde h$) in this limit. First let's redefine $\tilde q^2 = qy^2 u^{-6}$, pull out a factor of $(1-\tilde q)^{-1}$, and use (\ref{eq:rel2}) to rewrite (\ref{eq:simp}) as
\be
\lim_{i\rightarrow\infty} \sum_{\tilde h,k'} (-1)^{\tilde h+k'} c^{i, \tilde h,3i+5\tilde h-k'}y^{-2\tilde h} u^{\tilde h+k'} = \frac{i\th_1(6\nu-2z,2z)\eta(6\nu -2z)^3}{\th_4(6\nu -2z,2z)^2y^{1/2}u^{1/2}}\prod_{\ell, \ell'} \frac{1}{(1-y^\ell u^{\ell'})^{\tilde d^{(2,0)}(\ell, \ell')}}
\label{eq:preprealpha}
\ee
where we have defined
\be
\sum_{\ell, \ell'} \tilde d^{(2,0)}(\ell, \ell') y^\ell u^{\ell'} = g^{(2,0)}(y^{-2} u^6, y, u) - 1.
\label{eq:annoying}
\ee
The above formula can be unwieldy due to arbitrary high poles in $y$. For convenience let's define $\alpha \equiv uy^{-2}$ and, after unpacking the theta functions, rewrite (\ref{eq:preprealpha}) and (\ref{eq:annoying}) as 
\begin{align}
\lim_{i\rightarrow\infty}\sum_{\tilde h, k'} (-1)^{\tilde h+k'} c^{i, \tilde h, 3i+5\tilde h-k'}\a^{\tilde h} u^{k'} = &\(-\frac\a u\prod_{n=1}^{\infty}\frac{(1-\a^nu^{5n})^2(1-\a^{n-2}u^{5n-4})(1-\a^{n+1}u^{5n-1})}{(1-\a^{n-1}u^{5n-2})^2(1-\a^nu^{5n-3})^2} \) \nn \\ &~~~~~~~\prod_{\ell, \ell'}\frac{1}{(1-\alpha^{\ell}u^{\ell'})^{d^{(2,0)}(\ell,\ell')}},
\label{eq:qau}
\end{align}
where
\begin{align}
\sum_{\ell, \ell'} d^{(2,0)}(\ell, \ell') \a^\ell u^{\ell'} &= g^{(2,0)}(\alpha u^5, \alpha^{-1/2} u^{1/2}, u) - 1 \nn \\ 
&= \(\frac{1}{u(1-u^2)(1-u\a)(1-u^5\a)}\)\big((22u^3+u^5+u^7) \nn\\&~~~+ (1+21u^2-2u^3-21u^4-42u^5+u^6-2u^7-2u^8-2u^9)\a\nn\\&~~~+
(u^3-2u^6-u^7+23u^9+2u^{10}+u^{11})\a^2\big) \nn \\
&\equiv F^{(2,0)}(\a, u).
\label{eq:indiana}
\end{align}

Finally, we take the large $\tilde h$ limit of $\lim_{i\rightarrow\infty}(-1)^{\tilde h+k'}c^{i, \tilde h, 3i+5\tilde h-k'}$, corresponding to taking the left-moving conformal dimension of the state large. This is done by taking out a factor of $\frac{1}{(1-\a)}$ (from the $\ell=2,\ell'=0$ term in the product of (\ref{eq:qau})) and then using (\ref{eq:rel2}). Thus we get that
\be
\sum_{k'} \lim_{\tilde h\rightarrow\infty}\lim_{i\rightarrow\infty} (-1)^{\tilde h+k'} c^{i, \tilde h, 3i+5\tilde h-k'} u^{k'} = \(-\frac1{2u}\prod_{n=1}^{\infty}\frac{(1-u^{5n})^2(1-u^{5n-4})(1-u^{5n-1})}{(1-u^{5n-2})^2(1-u^{5n-3})^2} \) \prod_\ell \frac{1}{(1-u^\ell)^{D^{(2,0)}(\ell)}}
\label{eq:notyet}
\ee
where 
\begin{align}
\sum_\ell D^{(2,0)}(\ell) u^\ell &= F^{(2,0)}(1,u)-1 \nn\\
&=\frac{1+23u^2+44u^3+45u^4+25u^5+26u^6+3u^7+u^8}{u(u^5-1)}.
\label{eq:dl}
\end{align}
However, (\ref{eq:notyet}) and (\ref{eq:dl}) can be miraculously simplified to
\begin{align}
2\sum_{k'} \lim_{\tilde h\rightarrow\infty}&\lim_{i\rightarrow\infty} (-1)^{\tilde h+k'} c^{i, \tilde h, 3i+5\tilde h-k'} q^{k'} \nn\\ &= (1-q)^2(1-q^2)\prod_{n=1}^{\infty}\frac{1}{(1-q^{5n-4})^{25}(1-q^{5n-3})^{47}(1-q^{5n-2})^{47}(1-q^{5n-1})^{25}(1-q^{5n})^{24}}\nn\\
&= \frac{q(1-q)^2(1-q^2) G(\tau) H(\tau)^{23}}{\Delta(\tau)} \nn\\
&= 1 + 23q + 322q^2 + 3405 q^3 + 29833q^4 + 227126q^5 +  \mc{O}(q^6)
\label{eq:diffs}
\end{align}
where $G(\tau)$, $H(\tau)$, and $\Delta(\tau)$ are modular functions defined in Appendix \ref{app:thetachar}.

The generating function in (\ref{eq:diffs}) computes the large $\tilde h$ limit of $(-1)^{\tilde h+k'} \lim_{i\rightarrow\infty} c^{i, \tilde h, 3i+5\tilde h-k'}$. However, $\lim_{\tilde h\rightarrow\infty} \lim_{i\rightarrow\infty} c ^{i, \tilde h, 3i+5\tilde h-k'}$ depends on the parity of $\tilde h$. Thus this generating function is computing (up to a factor of $2$) the difference between the even and odd parity $\tilde h$ degeneracies (see Table \ref{tab:regulartable} for example). 

We can also compute the sum without much difficulty. Let's first rewrite (\ref{eq:qau}) as 
\begin{align}
\lim_{i\rightarrow\infty} \sum_{\tilde h,k'} c^{i, \tilde h, 3i+5\tilde h-k'} \a^{\tilde h} u^{k'} &= -\frac\a u\(\prod_{n=1}^{\infty} \frac{(1-\a^nu^{5n})^2(1-\a^{n-2}u^{5n-4})(1-\a^{n+1}u^{5n-1})}{(1+\a^{n-1}u^{5n-2})^2(1+\a^nu^{5n-3})^2}\) \nn \\ 
&~~~~~~~~~~~\(\prod_{\ell,\ell'} \frac{1}{(1-(-1)^{\ell+\ell'}\a^\ell u^{\ell'})^{d^{(2,0)}(\ell,\ell')}}\) \nn\\
&=  -\frac\a u\(\prod_{n=1}^{\infty} \frac{(1-\a^nu^{5n})^2(1-\a^{n-2}u^{5n-4})(1-\a^{n+1}u^{5n-1})}{(1+\a^{n-1}u^{5n-2})^2(1+\a^nu^{5n-3})^2}\) \nn \\ 
&~~~~~~~~~~~\(\prod_{\ell,\ell'} \frac{1}{(1-\a^\ell u^{\ell'})^{d^{(2,0)}_{\text{even}}(\ell,\ell')}(1+\a^\ell u^{\ell'})^{d^{(2,0)}_{\text{odd}}(\ell,\ell')}}\)
\end{align}
where
\begin{align}
\sum d^{(2,0)}_{\text{even}}(\ell, \ell') \a^\ell u^{\ell'} &= \frac{F^{(2,0)}(\a, u) + F^{(2,0)}(-\a, -u)}2\nn\\
\sum d^{(2,0)}_{\text{odd}}(\ell, \ell') \a^\ell u^{\ell'} &= \frac{F^{(2,0)}(\a, u) - F^{(2,0)}(-\a, -u)}2.
\end{align}
To extract the large $\tilde h$ limit, we again take out a factor of $\frac{1}{1-\a}$ and use (\ref{eq:rel2}). The final answer is 
\begin{align}
2\sum_{k'} \lim_{\tilde h\rightarrow\infty}&\lim_{i\rightarrow\infty} c^{i, \tilde h, 3i+5\tilde h-k'} q^{k'} \nn\\ 
&=q^{\frac65}(1-q^2)(1+q)^2 H(\tau)^{20} H(2\tau)^2\(\frac{R(2\tau)^{\frac{38}5}}{R(\tau)^{\frac{81}5}}\) \(\frac{\eta(4\tau)^{12}}{\eta(2\tau)^{36}}\)\prod_{n=1}^{\infty} \(\frac{(1+q^n)^{\frac{12n^2}5}}{(1-q^n)^{\frac{12n^2}5}}\) \nn\\
&= 1 + 23q + 322q^2 + 3405 q^3 + 29925q^4 + 229338q^5 + \mc{O}(q^6).
\label{eq:sumsk3}
\end{align}
This computes the sum (up to a factor of $2$) of the even and odd $\tilde h$ degeneracies. Again, see Table \ref{tab:regulartable} for examples.

Note we can also extract other information from (\ref{eq:qau}) and (\ref{eq:indiana}). For example we don't have to take the large $\tilde h$ limit; we can extract e.g. the $\tilde h=1$ limit easily from those equations; the final answer is (compare with Table \ref{tab:regulartable}):
\begin{align}
\lim_{i\rightarrow \infty} \sum_j c^{i, 1, 3i+4-2j} q^j &= (1-q)^2(1-q^2)(22q+23q^2+23q^3+2q^4+q^5+q^6)\frac{1}{\Delta(\tau)} \nn\\
&= 22 + 507 q + 6601 q^2 + 63044 q^3 + 489669 q^4 + 3265908 q^5 + \mc{O}(q^6).
\end{align}

\subsection{Maximal supergravity}

In this section we  do a similar analysis for the spectrum of multiparticle supergravity states arising from IIB compactification on AdS$_3\times S^3 \times T^4$. Here we will make use of characters of the contacted large $\mc{N}=4$ superconformal algebra, discussed in Appendix \ref{app:largen4}. As the derivations are extremely similar to those of the previous section, we will be brief.

\subsubsection{Half--BPS States}\label{sec:derivT4half}
In this section we derive the behavior of the degeneracies of 1/2--BPS states on Sym$^N(T^4)$ as $N\to \infty$. The full generating function of half-BPS states is given in equation (\ref{eq:hodgeT4}) has a character expansion given by 
\be
\sum_{N=0}^{\infty} \sum_{i,j=0}^{N-1} \check c_N^{i,j} p^N \check y_N^{[i]} \check u_N^{[j]} = \prod_{n=1}^{\infty} \frac{(1-p^nu^{-1})^2(1-p^nu)^2(1-p^ny^{-1})^2(1-p^ny)^2}{(1-p^n)^4(1-p^nu^{-1}y^{-1})(1-p^nu^{-1}y)(1-p^nuy^{-1})(1-p^nuy)},
\label{eq:halfbpst4}
\ee
where now $\check y_N^{[i]}$ is the $q^0$ term of the contracted large $\mc{N}=4$ Ramond character $\check\chi_{i;N}^{s,R}$ at central charge $6N$, i.e.
\be\label{eq:largecharbear}
\check y_N^{[i]} = (-1)^{N+i}\(\frac{y^{N-i}-y^{-(N-i)}}{y-y^{-1}}\)\(y^{-1}-2+y\).
\ee
To get the large $N$ limit, we plug (\ref{eq:largechar}) into (\ref{eq:halfbpst4}), redefine $\tilde p = puy,$ extract a factor of $(1-\tilde p)^{-1}$, and use (\ref{eq:rel2}) to get,
\begin{align}
&\lim_{N\rightarrow\infty} \sum_{i,j=0}^{N-1} (-1)^{i+j} \check c_N^{i,j} y^{-i} u^{-j} \nn\\
&= \frac{uy(1+u)(1+y)(uy-1)}{(1-u)^3(1-y)^3} \prod_{n=1}^{\infty} \frac{(1-u^{-n+1}y^{-n})^4(1-u^{-n}y^{-n+1})^4}{(1-u^{-n}y^{-n})^6(1-u^{-n-1}y^{-n+1})(1-u^{-n+1}y^{-n-1})}.
\end{align}
Next we take the large $i$ limit of by redefining $\tilde y = y^{-1}u^{-3}$, extracting a factor of $(1-\tilde y)^{-1}$, and using (\ref{eq:rel2}) to obtain
\be
\sum_k \lim_{i\rightarrow \infty}\lim_{N\rightarrow\infty} (-1)^{2k} \check c^{i, 3i-2k}_N u^{2k} = -\frac{u^2(1+u)(1-u^2)(1+u^3)}{(1-u)^3(1-u^3)^3(1-u^{-2})}\prod_{n=1}^{\infty} \frac{(1-u^{2n+1})^4(1-u^{2n-3})^4}{(1-u^{2n})^6 (1-u^{2n})(1-u^{2n+4})}. 
\ee
Simplifying this leads us to equation (\ref{eq:T4stab1}).

\subsubsection{Quarter--BPS States}\label{sec:derivT4quarter}
In this section we derive equation (\ref{eq:t4sumwoo}) which describes the large $N$, large left-moving charge limit of 1/4--BPS degeneracies presented in Table \ref{tab:regulartableT4}. 
The spectrum of multiparticle supergravity states is discussed in \S \ref{sec:spectrum}. The generating function of equation (\ref{eq:fullhegT4}) has a character decomposition given by
\be\label{eq:T4decomp}
\sum_{N=0}^\infty \sum_{i, \tilde h, j} p^N \check c_N^{i, \tilde h, j} \check \chi^{\ell, \text{NS}}_{i, \tilde h; N}(\t,z) \overline{\check \chi^{s, \text R}_{j;N}}(\nu) = \prod_{n>0, m, \ell, \ell'} \frac{1}{(1-p^n q^m y^\ell u^{\ell'})^{c_{\text{sugra}}^{(2,2)}(n,m,\ell,\ell')}}
\ee
where $c_{\text{sugra}}^{(2,2)}(n, m, \ell, \ell')$ is given in equation (\ref{eq:t4sugrakkearly}). 

At large central charge, the characters (\ref{eq:largechar}) simplify  via (\ref{eq:hawksatlargeC}). Thus we can rewrite equation (\ref{eq:T4decomp}) as
\begin{align}
\sum_{N,i,\tilde h,j}(-1)^{i+j} &p^Nq^{\frac i2+\tilde h} \check c_N^{i, \tilde h, j} (y^{i+1}-y^{-(i+1)})\(\frac{-i q^{3/8} \th_4(\t,z)^2 L^{\text{NS}}(\t,z)}{\th_1(\t,2z)\eta(\t)^6}\)\(\frac{(u^{N-j}-u^{-(N-j)})(u^{-1/2}-u^{1/2})}{u^{-1/2}+u^{1/2}}\) \nn \\
&=\frac{(1-p)^2}{(1-pu)(1-pu^{-1})}\prod_{n>1,m,\ell,\ell'}\frac{1}{(1-p^n q^m y^\ell u^{\ell'})^{c_{\text{sugra}}^{(2,2)}(n,m,\ell,\ell')}}.
\label{eq:t4allbps}
\end{align} 

We take the large $N$ limit by redefining $\tilde p \equiv pu$, extracting a factor of $(1-\tilde p)^{-1}$, and using (\ref{eq:rel2}) to get
\begin{align}
\sum_{i,\tilde h,j}(-1)^{i+j} &q^{\frac i2+\tilde h} \check c^{i, \tilde h, j} (y^{i+1}-y^{-(i+1)})u^{-j}=\(\frac{-i \th_1(\t,2z)\eta(\t)^6}{q^{3/8} \th_4(\t,z)^2 L^{\text{NS}}(\t,z)}\)\prod_{m,\ell,\ell'}\frac{1}{(1-q^m y^\ell u^{\ell'})^{c^{(2,2)}(m,\ell,\ell')}},
\end{align} 
where again we have dropped the subscript of $N$ on $\check c^{i, \tilde h, j}$ to indicate we have taken the limit of $N\to\infty$, and we define
\begin{align}
\sum_{m, \ell, \ell'} c^{(2,2)}&(m,\ell,\ell') q^m y^\ell u^{\ell'} = f^{(2,2)}(u^{-1}, q, y, u) - (1-2u^{-1}+u^{-2}) 
\equiv g^{(2,2)}(q, y, u).
\end{align}

Then  we take the large $i$ limit of $\check c^{i, \tilde h, 3i+5\tilde h-k'}$. Redefining $\tilde q \equiv q y^2 u^{-6}$, removing $(1-\tilde q^{1/2})^{-1}$, and using (\ref{eq:rel2}), we obtain
\begin{align}
\lim_{i\rightarrow\infty}\sum_{\tilde h,k'} (-1)^{\tilde h+k'} \check c^{i, \tilde h, 3i+5\tilde h-k'} y^{-2\tilde h} u^{\tilde h+k'}=\(\frac{-i \th_1(6\nu-2z,2z)\eta(6\nu-2z)^6}{y^{1/4} u^{9/4} \th_4(6\nu-2z,z)^2 L^{\text{NS}}(6\nu-2z,z)}\)\prod_{\ell,\ell'}\frac{1}{(1-y^{\ell} u^{\ell'})^{\tilde d^{(2,2)}(\ell,\ell')}}
\label{eq:notalphayet}
\end{align}
where we define
\be
\sum_{\ell, \ell'} \tilde d^{(2,2)}(\ell, \ell') y^\ell u^{\ell'} = g^{(2,2)}(y^{-2} u^6, y, u)-1.
\ee
Finally we rewrite (\ref{eq:notalphayet}) by defining $\alpha \equiv uy^{-2}$ to get
\begin{align}
\lim_{i\rightarrow\infty}\sum_{\tilde h,k'} (-1)^{\tilde h+k'} \check c^{i, \tilde h, 3i+5\tilde h-k'} \alpha^{h} u^{k'}&= -\frac{\a}u\(\prod_{n=1}^{\infty} \frac{(1-\a^nu^{5n})^6(1-\a^{n-2}u^{5n-4})(1-\a^{n+1}u^{5n-1})}{(1+\a^{n-1}u^{5n-2})^4(1+\a^nu^{5n-3})^4}\) \nn \\ 
&~~~~~~~~~~~\(\prod_{\ell,\ell'} \frac{1}{(1-(-1)^{\ell+\ell'}\a^\ell u^{\ell'})^{d^{(2,2)}(\ell,\ell')}}\),
\label{eq:alphayay}
\end{align}
where we define
\begin{align}
\sum_{\ell, \ell'} d^{(2,2)}(\ell, \ell') \a^\ell u^{\ell'} &= g^{(2,2)}(\alpha u^5, \alpha^{-1/2} u^{1/2}, u) - 1 
\equiv F^{(2,2)}(\a, u).
\label{eq:indianat4}
\end{align}

Unlike for $K3$, for $T^4$, the quantity $\lim_{\tilde h\rightarrow\infty} \lim_{i\rightarrow\infty} (-1)^{\tilde h+k'} \check c^{i,\tilde h, 3i+5\tilde h-k'}$ vanishes for all $k'$. However, we can instead compute $\lim_{\tilde h\rightarrow\infty} \lim_{i\rightarrow\infty} \check c^{i,\tilde h, 3i+5\tilde h-k'}$ using the same techniques as before. 

Let's rewrite (\ref{eq:alphayay}) as
\begin{align}
\lim_{i\rightarrow\infty}\sum_{\tilde h,k'} (-1)^{\tilde h+k'} \check c^{i, \tilde h, 3i+5\tilde h-k'} \alpha^{h} u^{k'}
&=  -\frac\a u\(\prod_{n=1}^{\infty} \frac{(1-\a^nu^{5n})^6(1-\a^{n-2}u^{5n-4})(1-\a^{n+1}u^{5n-1})}{(1+\a^{n-1}u^{5n-2})^4(1+\a^nu^{5n-3})^4}\) \nn \\ 
&~~~~~~~~~~~\(\prod_{\ell,\ell'} \frac{1}{(1-\a^\ell u^{\ell'})^{d^{(2,2)}_{\text{even}}(\ell,\ell')}(1+\a^\ell u^{\ell'})^{d^{(2,2)}_{\text{odd}}(\ell,\ell')}}\)
\end{align}
where
\begin{align}
\sum d^{(2,2)}_{\text{even}}(\ell, \ell') \a^\ell u^{\ell'} &= \frac{F^{(2,2)}(\a, u) + F^{(2,2)}(-\a, -u)}2\nn\\
\sum d^{(2,2)}_{\text{odd}}(\ell, \ell') \a^\ell u^{\ell'} &= \frac{F^{(2,2)}(\a, u) - F^{(2,2)}(-\a, -u)}2.
\end{align}
We extract the large $\tilde h$ limit, by taking out a factor of $\frac{1}{1-\a}$ and using (\ref{eq:rel2}) to obtain our result of (\ref{eq:t4sumwoo}). 

\section{Character and Decomposition Tables}
\label{sec:decomp}

In this Appendix, we give the character tables for the sporadic Mathieu groups $M_{11}$ and $M_{22}$, and various decompositions into $M_{11}$ and $M_{22}$ representations. We follow the  ATLAS notation \cite{atlas} for conjugacy classes and make use of the standard definitions $a_p := i \sqrt p; b_p := (-1 + i \sqrt p)/2$.

\begin{table}[!h]
\centering
\setstretch{1.2}
\begin{tabular}{|c|cccccccccccc|}
\hline
& 1A & 2A & 3A & 4A & 4B & 5A & 6A & 7A & 7B & 8A & 11A & 11B \\ \hline
$\chi_1$ & 1 & 1 & 1 & 1 & 1 & 1 & 1 & 1 & 1 & 1 & 1 & 1 \\
$\chi_2$ & 21 & 5 & 3 & 1 & 1 & 1 & $-1$ & 0 & 0 & $-1$ & $-1$ & $-1$ \\
$\chi_3$ & 45 & $-3$ & 0 & 1 & 1 & 0 & 0 & $b_7$ & $\bar b_7$ & $-1$ & 1 & 1 \\
$\chi_4$ & 45 & $-3$ & 0 & 1 & 1 & 0 & 0 & $\bar b_7$ & $b_7$ & $-1$ & 1 & 1 \\
$\chi_5$ & 55 & 7 & 1 & 3 & $-1$ & 0 & 1 & $-1$ & $-1$ & 1 & 0 & 0 \\
$\chi_6$ & 99 & 3 & 0 & 3 & $-1$ & $-1$ & 0 & 1 & 1 & $-1$ & 0 & 0 \\
$\chi_7$ & 154 & 10 & 1 & $-2$ & 2 & $-1$ & 1 & 0 & 0 & 0 & 0 & 0 \\
$\chi_8$ & 210 & 2 & 3 & $-2$ & $-2$ & 0 & $-1$ & 0 & 0 & 0 & 1 & 1 \\
$\chi_9$ & 231 & 7 & $-3$ & $-1$ & $-1$ & 1 & 1 & 0 & 0 & $-1$ & 0 & 0 \\
$\chi_{10}$ & 280 & $-8$ & 1 & 0 & 0 & 0 & 1 & 0 & 0 & 0 & $b_{11}$ &$\bar b_{11}$ \\
$\chi_{11}$ & 280 & $-8$ & 1 & 0 & 0 & 0 & 1 & 0 & 0 & 0 & $\bar b_{11}$ &$b_{11}$ \\
$\chi_{12}$ & 385 & 1 & $-2$ & 1 & 1 & 0 & $-2$ & 0 & 0 & 1 & 0 & 0 \\ \hline
$\pi_g$ & $1^{24}$ &  $1^82^8$ & $1^63^6$ &$1^42^24^4$ & $1^42^24^4$  & $1^45^4$ & $1^22^23^26^2$ & $1^37^3$ & $1^37^3$ & $1^22^14^18^2$ & $1^211^2$ & $1^211^2$ \\ \hline
\end{tabular}
\caption{Character table and Frame shapes for the group $M_{22}$.}
\label{table:m22char}
\end{table}

\begin{table}[!h]
\centering
\setstretch{1.2}
\begin{tabular}{|c|cccccccccc|}
\hline
& 1A & 2A & 3A & 4A & 5A & 6A & 8A & 8B & 11A & 11B \\ \hline
$\chi_1$ & 1 & 1 & 1 & 1 & 1 & 1 & 1 & 1 & 1 & 1 \\
$\chi_2$ & 10 & 2 & 1 & 2 & 0 & $-1$ & 0 & 0 & $-1$ & $-1$ \\
$\chi_3$ & 10 & $-2$ & 1 & 0 & 0 & 1 & $a_2$  & $\bar a_2$ & $-1$ & $-1$ \\
$\chi_4$ & 10 & $-2$ & 1 & 0 & 0 & 1 & $\bar a_2$ & $a_2$ & $-1$ & $-1$ \\
$\chi_5$ & 11 & 3 & 2 & $-1$ & 1 & 0 & $-1$ & $-1$ & 0 & 0 \\
$\chi_6$ & 16 & 0 & $-2$ & 0 & 1 & 0 & 0 & 0 & $b_{11}$ & $\bar b_{11}$ \\
$\chi_7$ & 16 & 0 & $-2$ & 0 & 1 & 0 & 0 & 0 & $\bar b_{11}$ & $b_{11}$ \\
$\chi_8$ & 44 & 4 & $-1$ & 0 & $-1$ & 1 & 0 & 0 & 0 & 0 \\
$\chi_9$ & 45 & $-3$ & 0 & 1 & 0 & 0 & $-1$ & $-1$ & 1 & 1 \\
$\chi_{10}$ & 55 & $-1$ & 1 & $-1$ & 0 & $-1$ & 1 & 1 & 0 & 0 \\ \hline
$\pi_g$ & $1^{24}$ & $1^8 2^8$ & $1^6 3^6$ & $2^44^4$ & $1^45^4$ & $1^22^23^26^2$ & $4^28^2$ & $4^2 8^2$ & $1^2 11^2$ & $1^2 11^2$ \\ \hline 
\end{tabular}
\caption{Character table and Frame shapes for the group $M_{11}$.}
\label{table:m11char}
\end{table}

\begin{sidewaystable}[!h]
\centering
\begin{tabular}{|c|cccccccccccc|}
\hline
& $\chi_1$& $\chi_2$& $\chi_3$& $\chi_4$& $\chi_5$& $\chi_6$& $\chi_7$& $\chi_8$& $\chi_9$& $\chi_{10}$& $\chi_{11}$& $\chi_{12}$ \\ \hline
$1q^0$& 1 & 0 & 0 & 0 & 0 & 0 & 0 & 0 & 0 & 0 & 0 & 0 \\
$23q^1$& 2 & 1 & 0 & 0 & 0 & 0 & 0 & 0 & 0 & 0 & 0 & 0 \\
$322q^2$& 8 & 5 & 0 & 0 & 1 & 0 & 1 & 0 & 0 & 0 & 0 & 0 \\
$3405q^3$& 22 & 20 & 0 & 0 & 6 & 1 & 7 & 4 & 1 & 0 & 0 & 1 \\
$29833q^4$& 72 & 77 & 0 & 0 & 33 & 10 & 43 & 26 & 14 & 5 & 5 & 19 \\
$227126q^5$& 199 & 273 & 4 & 4 & 151 & 72 & 223 & 173 & 126 & 75 & 75 & 171 \\
$1547673q^6$& 584 & 962 & 64 & 64 & 690 & 460 & 1150 & 1004 & 897 & 683 & 683 & 1278 \\
$9628056q^7$& 1613 & 3328 & 585 & 585 & 3051 & 2689 & 5740 & 5710 & 5637 & 5025 & 5025 & 8362 \\
$55464597q^8$ &4576 & 11813 & 4240 & 4240 & 13714 & 14921 & 28635 & 31168 & 32637 & 32120 & 32120 &
   50022 \\
$299037612q^9$ & 12807 & 43078 & 26476 & 26476 & 61883 & 78973 & 140856 & 164664 & 177372 & 186776 &
   186776 & 278575 \\ \hline
\end{tabular}
\caption{Decomposition of the expression given in (\ref{eq:ansdiffk3}) into $M_{22}$ representations.}
\label{table:diffm22}
\end{sidewaystable}

\clearpage

\begin{sidewaystable}[!h]
\centering
\begin{tabular}{|c|cccccccccccc|} 
\hline
& $\chi_1$& $\chi_2$& $\chi_3$& $\chi_4$& $\chi_5$& $\chi_6$& $\chi_7$& $\chi_8$& $\chi_9$& $\chi_{10}$& $\chi_{11}$& $\chi_{12}$ \\ \hline
$1q^0$ & 1 & 0 & 0 & 0 & 0 & 0 & 0 & 0 & 0 & 0 & 0 & 0 \\
$23q^1$ & 2 & 1 & 0 & 0 & 0 & 0 & 0 & 0 & 0 & 0 & 0 & 0 \\
$322q^2$ & 8 & 5 & 0 & 0 & 1 & 0 & 1 & 0 & 0 & 0 & 0 & 0 \\
$3405q^3$ & 22 & 20 & 0 & 0 & 6 & 1 & 7 & 4 & 1 & 0 & 0 & 1 \\
$29925q^4$ & 72 & 77 & 0 & 0 & 33 & 10 & 43 & 26 & 14 & 5 & 5 & 19 \\
$229338q^5$ & 199 & 273 & 4 & 4 & 151 & 72 & 223 & 173 & 126 & 75 & 75 & 171 \\
 $1579693q^6$ &584 & 962 & 64 & 64 & 690 & 460 & 1150 & 1004 & 897 & 683 & 683 & 1278 \\
$9976744q^7$ & 1613 & 3328 & 585 & 585 & 3051 & 2689 & 5740 & 5710 & 5637 & 5025 & 5025 & 8362 \\
$58605585q^8$ & 4576 & 11813 & 4240 & 4240 & 13714 & 14921 & 28635 & 31168 & 32637 & 32120 & 32120 &
   50022 \\
$323612308q^9$ & 12807 & 43078 & 26476 & 26476 & 61883 & 78973 & 140856 & 164664 & 177372 & 186776 &
   186776 & 278575 \\ \hline
\end{tabular}
\caption{Decomposition of the expression given in (\ref{eq:ansansdiffsk3}) into $M_{22}$ representations.}
\label{table:summ22}
\end{sidewaystable}

\clearpage

\begin{table}[!h]
\centering
\begin{tabular}{|c|cccccccccc|}
\hline
& $\chi_1$& $\chi_2$& $\chi_3$& $\chi_4$& $\chi_5$& $\chi_6$& $\chi_7$& $\chi_8$& $\chi_9$& $\chi_{10}$ \\ \hline
$1q^0$ & 1 & 0 & 0 & 0 & 0 & 0 & 0 & 0 & 0 & 0 \\
$23q^1$ & 1 & 0 & 0 & 0 & 2 & 0 & 0 & 0 & 0 & 0 \\
$322q^2$ & 6 & 3 & 0 & 0 & 9 & 0 & 0 & 3 & 0 & 1 \\
$3405q^3$ &  21 & 17 & 0 & 0 & 39 & 2 & 2 & 29 & 4 & 23 \\
$29833q^4$& 79 & 106 & 10 & 10 & 174 & 38 & 38 & 231 & 97 & 197 \\
$227126q^5$&  260 & 547 & 150 & 150 & 817 & 372 & 372 & 1570 & 979 & 1534 \\
 $1547673q^6$&  969 & 2966 & 1358 & 1358 & 3918 & 2826 & 2826 & 9926 & 7578 & 10520 \\
$9628056q^7$&  3638 & 15675 & 9826 & 9826 & 19341 & 18504 & 18504 & 58566 & 50257 & 65960 \\
$55464597q^8$ &  14607 & 81948 & 61536 & 61536 & 96456 & 109259 & 109259 & 326399 & 299783 & 381649 \\
$299037612q^9$ &  60672 & 415548 & 348624 & 348624 & 476324 & 596331 & 596331 & 1722786 & 1648636 &
   2064284 \\ \hline
\end{tabular}
\caption{Decomposition of the expression given in (\ref{eq:ansdiffk3}) into $M_{11}$ representations.} 
\label{table:diffm11}
\end{table} 

\begin{table}[!h]
\centering
\begin{tabular}{|c|cccccccccc|}
\hline
& $\chi_1$& $\chi_2$& $\chi_3$& $\chi_4$& $\chi_5$& $\chi_6$& $\chi_7$& $\chi_8$& $\chi_9$& $\chi_{10}$ \\ \hline
$1q^0$ &1 & 0 & 0 & 0 & 0 & 0 & 0 & 0 & 0 & 0 \\
$23q^1$ & 1 & 0 & 0 & 0 & 2 & 0 & 0 & 0 & 0 & 0 \\
$322q^2$ & 6 & 3 & 0 & 0 & 9 & 0 & 0 & 3 & 0 & 1 \\
$3405q^3$ & 21 & 17 & 0 & 0 & 39 & 2 & 2 & 29 & 4 & 23 \\
$29925q^4$ &   83 & 106 & 10 & 10 & 182 & 38 & 38 & 231 & 97 & 197 \\
$229338q^5$ & 288 & 563 & 150 & 150 & 857 & 372 & 372 & 1586 & 979 & 1550 \\
$1579693q^6$ & 1101 & 3098 & 1366 & 1366 & 4186 & 2858 & 2858 & 10170 & 7658 & 10740 \\
$9976744q^7$ &4214 & 16631 & 10050 & 10050 & 20865 & 19024 & 19024 & 60978 & 51681 & 68332 \\
$58605585q^8$ &  17123 & 88360 & 64240 & 64240 & 105420 & 114779 & 114779 & 346631 & 314787 & 403097 \\
$323612308q^9$ &  71752 & 457084 & 373552 & 373552 & 528908 & 642723 & 642723 & 1872678 & 1775728 &
   2232872 \\ \hline
\end{tabular}
\caption{Decomposition of the expression given in (\ref{eq:ansansdiffsk3}) into $M_{11}$ representations.}
\label{table:summ11}
\end{table}

\clearpage

\begin{table}[h]
\begin{center}
\begin{tabular}{| c | c  c  c  c  c  c  c  c  c  c  c  c |}
\hline
Degeneracy & $\chi_1$& $\chi_2$& $\chi_3$& $\chi_4$& $\chi_5$& $\chi_6$& $\chi_7$& $\chi_8$& $\chi_9$& $\chi_{10}$& $\chi_{11}$& $\chi_{12}$ \\ \hline210 & 0 & 0 & 0 & 0 & 0 & 0 & 0 & 1 & 0 & 0 & 0 & 0 \\ 
21 & 0 & 1 & 0 & 0 & 0 & 0 & 0 & 0 & 0 & 0 & 0 & 0 \\ 
3542 & 1 & 3 & 0 & 0 & 2 & 1 & 3 & 3 & 2 & 1 & 1 & 3 \\ 
484 & 2& 3& 0& 0& 1& 0& 1& 1& 0& 0& 0& 0\\
22 & 1& 1& 0& 0& 0& 0& 0& 0& 0& 0& 0& 0\\ 
21 &0& 1& 0& 0& 0& 0& 0& 0& 0& 0& 0& 0\\ 
36961 & 4& 13& 1& 1& 12& 10& 22& 26& 21& 18& 18& 29\\
6281 & 6& 12& 0& 0& 6& 2& 8& 7& 3& 1& 1& 4\\
506 & 3& 4& 0& 0& 1& 0& 1& 1& 0& 0& 0& 0\\ 
22 & 1& 1& 0& 0& 0& 0& 0& 0& 0& 0& 0& 0\\ 
231 & 1& 1& 0& 0& 1& 0& 1& 0& 0& 0& 0& 0\\ 
2660 & 0& 0& 0& 0& 0& 0& 0& 2& 0& 4& 4& 0\\ 
21526 & 5& 10& 1& 1& 10& 7& 17& 10& 15& 8& 8& 19\\ 
420 & 0& 0& 0& 0& 0& 0& 0& 2& 0& 0& 0& 0\\ 
3796 & 4& 5& 0& 0& 3& 1& 4& 3& 2& 1& 1& 3\\ 
275 & 3& 3& 0& 0& 1& 0& 1& 0& 0& 0& 0& 0\\ 
1 & 1& 0& 0& 0& 0& 0& 0& 0& 0& 0& 0& 0\\ \hline

\end{tabular}
\end{center}
\caption{Decomposition of all terms in (\ref{eq:sushi}) into $M_{22}$ representations.}
\label{tab:toro}
\end{table}

\bibliographystyle{JHEP}
\bibliography{refs}

\providecommand{\href}[2]{#2}\begingroup\raggedright\begin{thebibliography}{10}

\bibitem{Strominger:1996sh}
A.~Strominger and C.~Vafa, \emph{{Microscopic origin of the Bekenstein-Hawking
  entropy}}, \href{http://dx.doi.org/10.1016/0370-2693(96)00345-0}{\emph{Phys.
  Lett.} {\bf B379} (1996) 99--104},
  [\href{http://arxiv.org/abs/hep-th/9601029}{{\tt hep-th/9601029}}].

\bibitem{Maldacena:1997re}
J.~M. Maldacena, \emph{{The Large N limit of superconformal field theories and
  supergravity}}, \href{http://dx.doi.org/10.1023/A:1026654312961}{\emph{Int.
  J. Theor. Phys.} {\bf 38} (1999) 1113--1133},
  [\href{http://arxiv.org/abs/hep-th/9711200}{{\tt hep-th/9711200}}].

\bibitem{Maldacena:1999bp}
J.~M. Maldacena, G.~W. Moore and A.~Strominger, \emph{{Counting BPS black holes
  in toroidal Type II string theory}},
  \href{http://arxiv.org/abs/hep-th/9903163}{{\tt hep-th/9903163}}.

\bibitem{Kachru:2016igs}
S.~Kachru and A.~Tripathy, \emph{{The Hodge-elliptic genus, spinning BPS
  states, and black holes}},
  \href{http://dx.doi.org/10.1007/s00220-017-2910-1}{\emph{Commun. Math. Phys.}
  {\bf 355} (2017) 245--259}, [\href{http://arxiv.org/abs/1609.02158}{{\tt
  1609.02158}}].

\bibitem{Benjamin:2016pil}
N.~Benjamin, \emph{{A Refined Count of BPS States in the D1/D5 System}},
  \href{http://dx.doi.org/10.1007/JHEP06(2017)028}{\emph{JHEP} {\bf 06} (2017)
  028}, [\href{http://arxiv.org/abs/1610.07607}{{\tt 1610.07607}}].

\bibitem{Benjamin:2017xen}
N.~Benjamin, S.~Kachru and A.~Tripathy, \emph{{Counting spinning dyons in
  maximal supergravity: The Hodge-elliptic genus for tori}},
  \href{http://dx.doi.org/10.1007/s11005-017-0981-8}{\emph{Lett. Math. Phys.}
  {\bf 107} (2017) 2081--2092}, [\href{http://arxiv.org/abs/1704.05423}{{\tt
  1704.05423}}].

\bibitem{Eguchi:2010ej}
T.~Eguchi, H.~Ooguri and Y.~Tachikawa, \emph{{Notes on the K3 Surface and the
  Mathieu group $M_{24}$}},
  \href{http://dx.doi.org/10.1080/10586458.2011.544585}{\emph{Exper. Math.}
  {\bf 20} (2011) 91--96}, [\href{http://arxiv.org/abs/1004.0956}{{\tt
  1004.0956}}].

\bibitem{deBoer:1998kjm}
J.~de~Boer, \emph{{Six-dimensional supergravity on S**3 x AdS(3) and 2-D
  conformal field theory}},
  \href{http://dx.doi.org/10.1016/S0550-3213(99)00160-1}{\emph{Nucl. Phys.}
  {\bf B548} (1999) 139--166}, [\href{http://arxiv.org/abs/hep-th/9806104}{{\tt
  hep-th/9806104}}].

\bibitem{deBoer:1998us}
J.~de~Boer, \emph{{Large N elliptic genus and AdS / CFT correspondence}},
  \href{http://dx.doi.org/10.1088/1126-6708/1999/05/017}{\emph{JHEP} {\bf 05}
  (1999) 017}, [\href{http://arxiv.org/abs/hep-th/9812240}{{\tt
  hep-th/9812240}}].

\bibitem{Dijkgraaf:1996xw}
R.~Dijkgraaf, G.~W. Moore, E.~P. Verlinde and H.~L. Verlinde, \emph{{Elliptic
  genera of symmetric products and second quantized strings}},
  \href{http://dx.doi.org/10.1007/s002200050087}{\emph{Commun. Math. Phys.}
  {\bf 185} (1997) 197--209}, [\href{http://arxiv.org/abs/hep-th/9608096}{{\tt
  hep-th/9608096}}].

\bibitem{Dijkgraaf:2000fq}
R.~Dijkgraaf, J.~M. Maldacena, G.~W. Moore and E.~P. Verlinde, \emph{{A Black
  hole Farey tail}},  \href{http://arxiv.org/abs/hep-th/0005003}{{\tt
  hep-th/0005003}}.

\bibitem{Maldacena:1998bw}
J.~M. Maldacena and A.~Strominger, \emph{{AdS(3) black holes and a stringy
  exclusion principle}},
  \href{http://dx.doi.org/10.1088/1126-6708/1998/12/005}{\emph{JHEP} {\bf 12}
  (1998) 005}, [\href{http://arxiv.org/abs/hep-th/9804085}{{\tt
  hep-th/9804085}}].

\bibitem{Sevrin:1988ew}
A.~Sevrin, W.~Troost and A.~Van~Proeyen, \emph{{Superconformal Algebras in
  Two-Dimensions with N=4}},
  \href{http://dx.doi.org/10.1016/0370-2693(88)90645-4}{\emph{Phys. Lett.} {\bf
  B208} (1988) 447--450}.

\bibitem{Gottsche}
L.~G{\"o}ttsche and W.~Soergel, \emph{{Perverse Sheaves and the Cohomology of
  Hilbert Schemes of Smooth Algebraic Surfaces}}, {\emph{Math. Ann.} {\bf 296}
  (1993) 235--245}.

\bibitem{Narain:1986am}
K.~S. Narain, M.~H. Sarmadi and E.~Witten, \emph{{A Note on Toroidal
  Compactification of Heterotic String Theory}},
  \href{http://dx.doi.org/10.1016/0550-3213(87)90001-0}{\emph{Nucl. Phys.} {\bf
  B279} (1987) 369--379}.

\bibitem{Aspinwall:1994rg}
P.~S. Aspinwall and D.~R. Morrison, \emph{{String theory on K3 surfaces}},
  \href{http://arxiv.org/abs/hep-th/9404151}{{\tt hep-th/9404151}}.

\bibitem{Nahm:1999ps}
W.~Nahm and K.~Wendland, \emph{{A Hiker's guide to K3: Aspects of N=(4,4)
  superconformal field theory with central charge c = 6}},
  \href{http://dx.doi.org/10.1007/PL00005548}{\emph{Commun. Math. Phys.} {\bf
  216} (2001) 85--138}, [\href{http://arxiv.org/abs/hep-th/9912067}{{\tt
  hep-th/9912067}}].

\bibitem{Gaberdiel:2011fg}
M.~R. Gaberdiel, S.~Hohenegger and R.~Volpato, \emph{{Symmetries of K3 sigma
  models}}, \href{http://dx.doi.org/10.4310/CNTP.2012.v6.n1.a1}{\emph{Commun.
  Num. Theor. Phys.} {\bf 6} (2012) 1--50},
  [\href{http://arxiv.org/abs/1106.4315}{{\tt 1106.4315}}].

\bibitem{Cheng:2016org}
M.~C.~N. Cheng, S.~M. Harrison, R.~Volpato and M.~Zimet, \emph{{K3 String
  Theory, Lattices and Moonshine}},
  \href{http://arxiv.org/abs/1612.04404}{{\tt 1612.04404}}.

\bibitem{Volpato:2014zla}
R.~Volpato, \emph{{On symmetries of $\mathcal{N}=(4,4)$ sigma models on
  $T^4$}}, \href{http://dx.doi.org/10.1007/JHEP08(2014)094}{\emph{JHEP} {\bf
  08} (2014) 094}, [\href{http://arxiv.org/abs/1403.2410}{{\tt 1403.2410}}].

\bibitem{Paquette:2017gmb}
N.~M. Paquette, R.~Volpato and M.~Zimet, \emph{{No More Walls! A Tale of
  Modularity, Symmetry, and Wall Crossing for 1/4 BPS Dyons}},
  \href{http://dx.doi.org/10.1007/JHEP05(2017)047}{\emph{JHEP} {\bf 05} (2017)
  047}, [\href{http://arxiv.org/abs/1702.05095}{{\tt 1702.05095}}].

\bibitem{Dijkgraaf:1998gf}
R.~Dijkgraaf, \emph{{Instanton strings and hyperKahler geometry}},
  \href{http://dx.doi.org/10.1016/S0550-3213(98)00869-4}{\emph{Nucl. Phys.}
  {\bf B543} (1999) 545--571}, [\href{http://arxiv.org/abs/hep-th/9810210}{{\tt
  hep-th/9810210}}].

\bibitem{Seiberg:1999xz}
N.~Seiberg and E.~Witten, \emph{{The D1 / D5 system and singular CFT}},
  \href{http://dx.doi.org/10.1088/1126-6708/1999/04/017}{\emph{JHEP} {\bf 04}
  (1999) 017}, [\href{http://arxiv.org/abs/hep-th/9903224}{{\tt
  hep-th/9903224}}].

\bibitem{Cheng:2010pq}
M.~C.~N. Cheng, \emph{{K3 Surfaces, N=4 Dyons, and the Mathieu Group M24}},
  \href{http://dx.doi.org/10.4310/CNTP.2010.v4.n4.a2}{\emph{Commun. Num. Theor.
  Phys.} {\bf 4} (2010) 623--658}, [\href{http://arxiv.org/abs/1005.5415}{{\tt
  1005.5415}}].

\bibitem{Cheng:2015kha}
M.~C.~N. Cheng, J.~F.~R. Duncan, S.~M. Harrison and S.~Kachru,
  \emph{{Equivariant K3 Invariants}},
  \href{http://dx.doi.org/10.4310/CNTP.2017.v11.n1.a2}{\emph{Commun. Num.
  Theor. Phys.} {\bf 11} (2017) 41--72},
  [\href{http://arxiv.org/abs/1508.02047}{{\tt 1508.02047}}].

\bibitem{Yellow}
P.~Di~Francesco, P.~Mathieu and D.~Senechal, \emph{Conformal Field Theory}.
\newblock Springer, 1997.

\bibitem{Nakajima}
H.~Nakajima, \emph{Heisenberg algebra and hilbert schemes of points on
  projective surfaces}, {\emph{Annals of Mathematics} {\bf 145} (1997)
  379--388}, [\href{http://arxiv.org/abs/alg-geom/9507012}{{\tt
  alg-geom/9507012}}].

\bibitem{Raju:2007uj}
S.~Raju, \emph{{Counting giant gravitons in AdS(3)}},
  \href{http://dx.doi.org/10.1103/PhysRevD.77.046012}{\emph{Phys. Rev.} {\bf
  D77} (2008) 046012}, [\href{http://arxiv.org/abs/0709.1171}{{\tt
  0709.1171}}].

\bibitem{Emparan:2006mm}
R.~Emparan and H.~S. Reall, \emph{{Black Rings}},
  \href{http://dx.doi.org/10.1088/0264-9381/23/20/R01}{\emph{Class. Quant.
  Grav.} {\bf 23} (2006) R169},
  [\href{http://arxiv.org/abs/hep-th/0608012}{{\tt hep-th/0608012}}].

\bibitem{Gukov:2004ym}
S.~Gukov, E.~Martinec, G.~W. Moore and A.~Strominger, \emph{{The Search for a
  holographic dual to AdS(3) x S**3 x S**3 x S**1}},
  \href{http://dx.doi.org/10.4310/ATMP.2005.v9.n3.a3,
  10.1142/9789812775344_0035}{\emph{Adv. Theor. Math. Phys.} {\bf 9} (2005)
  435--525}, [\href{http://arxiv.org/abs/hep-th/0403090}{{\tt
  hep-th/0403090}}].

\bibitem{Eberhardt:2017fsi}
L.~Eberhardt, M.~R. Gaberdiel, R.~Gopakumar and W.~Li, \emph{{BPS spectrum on
  AdS$_3\times $S$^3 \times $S$^3 \times $S$^1$}},
  \href{http://dx.doi.org/10.1007/JHEP03(2017)124}{\emph{JHEP} {\bf 03} (2017)
  124}, [\href{http://arxiv.org/abs/1701.03552}{{\tt 1701.03552}}].

\bibitem{Eberhardt:2017pty}
L.~Eberhardt, M.~R. Gaberdiel and W.~Li, \emph{{A holographic dual for string
  theory on AdS$_{3}$×S$^{3}$×S$^{3}$×S$^{1}$}},
  \href{http://dx.doi.org/10.1007/JHEP08(2017)111}{\emph{JHEP} {\bf 08} (2017)
  111}, [\href{http://arxiv.org/abs/1707.02705}{{\tt 1707.02705}}].

\bibitem{Maldacena:1997de}
J.~M. Maldacena, A.~Strominger and E.~Witten, \emph{{Black hole entropy in M
  theory}}, \href{http://dx.doi.org/10.1088/1126-6708/1997/12/002}{\emph{JHEP}
  {\bf 12} (1997) 002}, [\href{http://arxiv.org/abs/hep-th/9711053}{{\tt
  hep-th/9711053}}].

\bibitem{Wong:2017wfe}
K.~Wong, \emph{{Quarter-BPS states in orbifold sigma models with ADE
  singularities}}, \href{http://dx.doi.org/10.1007/JHEP06(2017)116}{\emph{JHEP}
  {\bf 06} (2017) 116}, [\href{http://arxiv.org/abs/1704.02926}{{\tt
  1704.02926}}].

\bibitem{Cheng:2014zpa}
M.~C.~N. Cheng and S.~Harrison, \emph{{Umbral Moonshine and K3 Surfaces}},
  \href{http://dx.doi.org/10.1007/s00220-015-2398-5}{\emph{Commun. Math. Phys.}
  {\bf 339} (2015) 221--261}, [\href{http://arxiv.org/abs/1406.0619}{{\tt
  1406.0619}}].

\bibitem{Cheng:2013wca}
M.~C.~N. Cheng, J.~F.~R. Duncan and J.~A. Harvey, \emph{{Umbral Moonshine and
  the Niemeier Lattices}},  \href{http://arxiv.org/abs/1307.5793}{{\tt
  1307.5793}}.

\bibitem{Duncan:2014eha}
J.~F.~R. Duncan and S.~Mack-Crane, \emph{{The Moonshine Module for Conway's
  Group}}, \href{http://dx.doi.org/10.1017/fms.2015.7}{\emph{SIGMA} {\bf 3}
  (2015) e10}, [\href{http://arxiv.org/abs/1409.3829}{{\tt 1409.3829}}].

\bibitem{Cheng:2014owa}
M.~C.~N. Cheng, X.~Dong, J.~F.~R. Duncan, S.~Harrison, S.~Kachru and T.~Wrase,
  \emph{{Mock Modular Mathieu Moonshine Modules}},
  \href{http://arxiv.org/abs/1406.5502}{{\tt 1406.5502}}.

\bibitem{Dabholkar:2012nd}
A.~Dabholkar, S.~Murthy and D.~Zagier, \emph{{Quantum Black Holes, Wall
  Crossing, and Mock Modular Forms}},
  \href{http://arxiv.org/abs/1208.4074}{{\tt 1208.4074}}.

\bibitem{RR1}
G.~E. Andrews and B.~Berndt, \emph{Ramanujan's Lost Notebook Part I}.
\newblock Springer, 2005.

\bibitem{RR2}
B.~Berndt, \emph{Ramanujan's Notebooks Part V}.
\newblock Springer, 1998.

\bibitem{Eguchi:1987sm}
T.~Eguchi and A.~Taormina, \emph{{Unitary Representations of $N=4$
  Superconformal Algebra}},
  \href{http://dx.doi.org/10.1016/0370-2693(87)91679-0}{\emph{Phys. Lett.} {\bf
  B196} (1987) 75}.

\bibitem{Eguchi:1987wf}
T.~Eguchi and A.~Taormina, \emph{{Character Formulas for the $N=4$
  Superconformal Algebra}},
  \href{http://dx.doi.org/10.1016/0370-2693(88)90778-2}{\emph{Phys. Lett.} {\bf
  B200} (1988) 315}.

\bibitem{Petersen:1989zz}
J.~L. Petersen and A.~Taormina, \emph{{Characters of the $N=4$ Superconformal
  Algebra With Two Central Extensions}},
  \href{http://dx.doi.org/10.1016/0550-3213(90)90084-Q}{\emph{Nucl. Phys.} {\bf
  B331} (1990) 556--572}.

\bibitem{Petersen:1989pp}
J.~L. Petersen and A.~Taormina, \emph{{Characters of the $N=4$ Superconformal
  Algebra With Two Central Extensions: 2. Massless Representations}},
  \href{http://dx.doi.org/10.1016/0550-3213(90)90141-Y}{\emph{Nucl. Phys.} {\bf
  B333} (1990) 833--854}.

\bibitem{atlas}
J.~H. Conway, R.~T. Curtis, S.~P. Norton, R.~A. Parker and R.~A. Wilson,
  \emph{Atlas of Finite Groups}.
\newblock Oxford University Press, 1985.

\end{thebibliography}\endgroup
\end{document}